\documentclass[reprint,showpacs,preprintnumbers,amsmath,amssymb, citeautoscript,prl,superscriptaddress]{revtex4-1}
\usepackage{graphicx,wasysym}
\usepackage{dcolumn}
\usepackage{bm}
\setcitestyle{super}

\usepackage{hyperref}
\usepackage{color}
\usepackage{algorithm}
\usepackage{algpseudocode}

\begin{document}
\title{Nonequilibrium design strategies for functional colloidal assemblies}

\preprint{APS/123-QED}
\author{Avishek Das}
\affiliation{Department of Chemistry, University of California, Berkeley, CA, 94720, USA \looseness=-1}
\author{David T. Limmer}
\email{dlimmer@berkeley.edu}
\affiliation{Department of Chemistry, University of California, Berkeley, CA, 94720, USA \looseness=-1}
\affiliation{Chemical Sciences Division, LBNL, Berkeley, CA, 94720, USA \looseness=-1}
\affiliation{Material Sciences Division, LBNL, Berkeley, CA, 94720, USA \looseness=-1}
\affiliation{Kavli Energy NanoSciences Institute, University of California, Berkeley, CA, 94720, USA \looseness=-1}

\date{\today}

\begin{abstract}
We use a nonequilibrium variational principle to optimize the steady-state, shear-induced interconversion of self-assembled nanoclusters of DNA-coated colloids. Employing this principle within a stochastic optimization algorithm allows us to discover design strategies for functional materials. We find that far-from-equilibrium shear flow can significantly enhance the flux between specific colloidal states by decoupling trade-offs between stability and reactivity required by systems in equilibrium. For isolated nanoclusters, we find nonequilibrium strategies for amplifying transition rates by coupling a given reaction coordinate to the background shear flow. 
We also find that shear flow can be made to selectively break detailed balance and maximize probability currents by coupling orientational degrees of freedom to conformational transitions. 
For a microphase consisting of many nanoclusters, we study the flux of colloids hopping between clusters. We find that a shear flow can amplify the flux without a proportional compromise on the microphase structure. This approach provides a general means of uncovering design principles for nanoscale, autonomous, functional materials driven far from equilibrium. 
\end{abstract}

\maketitle


\thispagestyle{empty}
\appendix

Self-assembly is a dynamic process and its optimal design involves simultaneously stabilizing a target structure while preserving a high rate of conversion into it from a fixed set of precursors \cite{whitesides2002self,jacobs2015rational,fullerton2016optimising,coropceanu2022self}.
In and near thermal equilibrium, fundamental trade-offs exist between these two design criteria making them difficult to fulfill concurrently \cite{whitelam2015statistical}.
For example, strong, specific interactions are needed to stabilize select structures over others, but strong interactions result in the slow relaxation of defects, and highly specific interactions can stifle growth \cite{whitelam2009role}. 
%
It is known that far-from-equilibrium conditions may generally accelerate rare reactive events \cite{kuznets2021dissipation,das2022direct}, and this has been leveraged in simple models of self-assembly \cite{whitelam2018strong,nguyen2021organization}.
However, the extent to which nonequilibrium conditions can aid the design of self-assembling systems by breaking  constraints between the factors that stabilize structures and those that aid in the dynamics of their formation is largely unknown. Moreover, there is currently a lack of general principles to inform strategies employing nonequilibrium conditions to assemble materials with unique structure or function, beyond what is possible in equilibrium. Specific examples in colloidal systems however point to the potential utility of such strategies \cite{charbonneau2007phase,chremos2010ultra,stenhammar2013continuum}. Here we present a framework for learning the optimal design principles for a general nonequilibrium self-assembling system, and demonstrate how the continuous injection of energy significantly expands the available design space of functional materials assembly.  

Functional materials cycle between multiple conformations in order to achieve a dynamical task. Any functional structure like an enzyme or a molecular motor must not only be stable, but also necessarily able to disassemble and reassemble in a dynamic nonequilibrium steady-state \cite{wang2009tuning,atkin2017disassembly}. Such materials typically form internal reaction networks of disassembled intermediates with multiple nonzero reactive fluxes, driven by external forces that are autonomous \cite{grzybowski2017dynamic,murugan2012speed,knoch2017nonequilibrium,albaugh2022simulating}. Autonomous driving is time-independent, energetically cheap to implement and acts nonspecifically on the entire system \cite{seifert2012stochastic}.
Functional materials efficiently transduce chemical and mechanical energy by optimally coupling the external dissipative forces to specific edges in their internal reaction networks. This may modify reactive fluxes and break detailed balance in those reaction coordinates that are then coupled to their output work. Thus, optimizing the dynamics of self-assembly for probability fluxes in and out of a target structure is the path to designing functional materials. However, a framework for doing so is complicated as typical relationships concerning stability and response valid in equilibrium no longer hold. 

Using a nonequilibrium variational framework, we develop a route to discover design rules for driven self-assembly. This framework uses insights from large deviation theory and stochastic thermodynamics \cite{das2021variational,das2019variational}. In particular, we probe the dynamics and outcome of self-assembly by analyzing order parameters over trajectories from molecular dynamics simulations \cite{hagan2006dynamic,newton2015rotational,jack2007fluctuation}.
Previous work on optimizing the reactivity of self-assembled colloidal clusters have used Kramers approximations valid only at equilibrium \cite{goodrich2020self} or have looked at single transient realizations of reactive barrier crossing \cite{miskin2016turning}. The design of optimal disassembly and reassembly, \textit{i.e.}, nonzero probability fluxes and currents in and out of functional structures in nonequilibrium steady-states, remains unexplored. Employing a variational principle, we adapt a previous algorithm used to study the optimal design of passive DNA-coated colloids, to design instead for enhanced reactive fluxes. We study the effectiveness of this approach in isolated colloidal nanoclusters and in a microphase of many nanoclusters, both in a sheared nonequilibrium steady-state. Using the inverse design algorithm and rationalizing the discovered designs in parameter space, we find nonequilibrium strategies for breaking trade-offs between structure and dynamics.


\subsection*{Model of DNA coated colloids}
\begin{figure*}[t]
\centering
\includegraphics[trim={3cm 0 0 0},width=17.6cm]{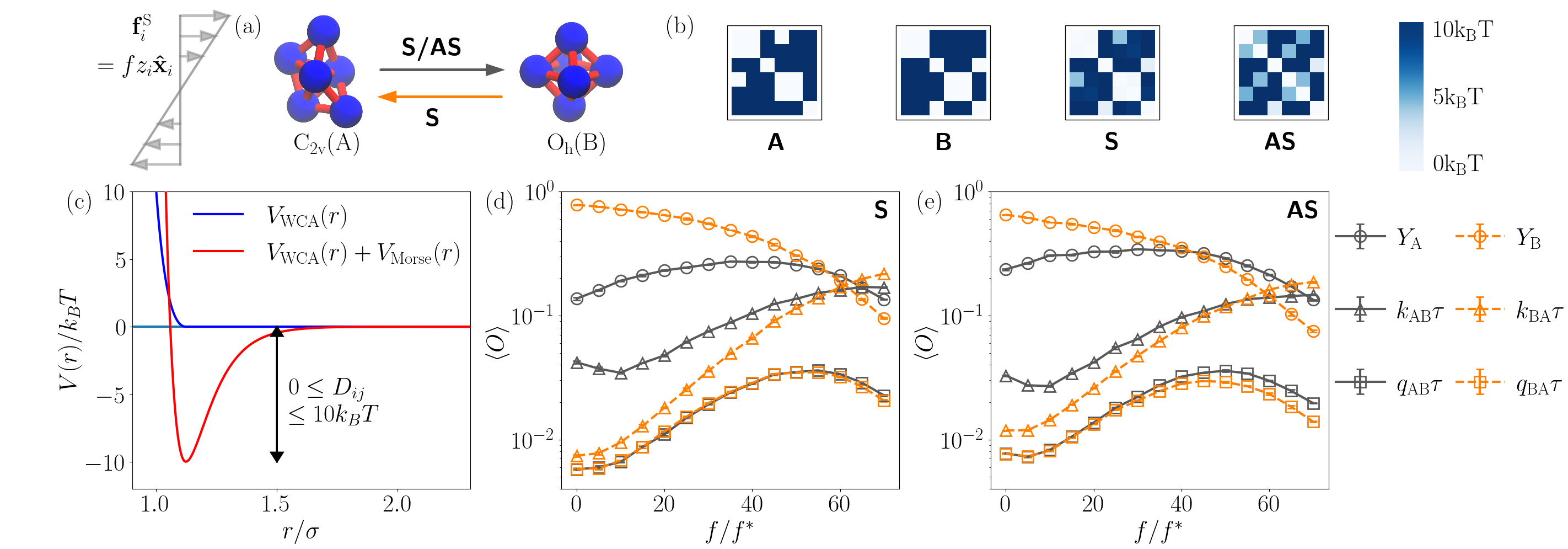}
\caption{Optimal probability flux between polytetrahedron and octahedron. (a) Schematic of isomerization reaction. Volume exclusion due to the colloids are shown in blue, and tunable pairwise bonds are shown in red. (b) Optimal $\textit{alphabets}$ for high yield and high flux. (c) Pairwise interaction potential.  (d) and (e) Average values of observable $\langle O\rangle\in\{Y,k\tau,q\tau\}$ with respect to shear, keeping the \textit{alphabet}s fixed at $\mathbf{S}$ and $\mathbf{AS}$.}
\label{fig:fig1}
\end{figure*}

We have modeled the self-assembly of DNA-labeled colloidal nanoparticles via overdamped Langevin dynamics simulations of $N$ particles, interacting pairwise with a coarse-grained potential $V(r)$, with $r$ being the center-to-center distance between two colloids, see Fig.~\ref{fig:fig1}c. The potential consists of a volume-exclusion repulsion modeled with a Weeks-Chandler-Andersen (WCA) potential \cite{weeks1971role} with an exclusion diameter of $\sigma$, and selective short-range attractive interactions modeled with a Morse potential with amplitude $D_{ij}$ between pairwise particle indices $i$ and $j$. Such a potential has been demonstrated previously to accurately represent selective attractions between two DNA-labeled surfaces \cite{rogers2011direct,manoharan2015colloidal}. Self-assembly with this model can be designed to yield nanoclusters of pre-specified sizes and symmetry, by tuning $D_{ij}$ with respect to $k_{\mathrm{B}}T$, or Boltzmann's constant times temperature \cite{meng2010free,zeravcic2014size,das2021variational}. We consider this self-assembling system evolving within a nonequilibrium steady-state under a shear flow with a constant shear flow rate $f$. The shear flow displaces the particles by competing with the natural diffusive velocity gradient scale for the system, $f^{*}=k_{\mathrm{B}}T/\sigma^{2}$. Given the mechanical driving force of shear performs dissipative work on the system, it may change the relative stability of configurations as well as the rate of interconversions between them. 
We investigate whether optimal design principles can be found that can tune the dynamics of self-assembly independently from configurational probabilities for pre-specified target order parameters.

\subsection*{Equilibrium constraints on self-assembly dynamics}
Thermal equilibrium constrains the dynamics of rates of interconversion between stable self-assembled structures to the yields of those structures. Both quantities can be viewed as trajectory-averaged quantities. 
We denote a simulated trajectory of particle positions $\mathbf{r}^{N}(t)$ as a function of time $t$, by $X=\{\mathbf{r}^{N}(0),\mathbf{r}^{N}(\delta t),\dots,\mathbf{r}^{N}(t_{f})\}$, with $t_{f}$ being the duration of the trajectory and $\delta t$ the simulation time step. We take $t_{f}$ much longer than the free colloidal diffusion timescale $t^{*}=\gamma\sigma^{2}/k_{\mathrm{B}}T$, where $\gamma$ is the friction coefficient for the free colloids in solution, such that the colloids relax into a steady-state and trajectory averages of order parameters are independent of $t_{f}$. The expected yield, $Y_\mathrm{A}$, of any target structure A over this trajectory can be computed from a trajectory average of the indicator function, 
\begin{equation}
Y_{\mathrm{A}}=\frac{1}{t_{f}}\int_{0}^{t_{f}}\mathbf{1}_{\mathrm{A}}[r^{N}(t)] dt\, ,
\end{equation}
where $\mathbf{1}_{\mathrm{A}}[r^{N}(t)]$ returns 1 when the system is in state A and is 0 otherwise. For computing the interconversion rates, we first define the probability flux from state A to state B, $q_{\mathrm{AB}}$. The steady state flux can be evaluated from a two-time correlation function between the indicator functions for state A and B over a lag-time $\tau$,
\begin{equation}
q_{\mathrm{AB}}=\frac{1}{\tau(t_{f}-\tau)}\int_{\tau}^{t_{f}}\mathbf{1}_{\mathrm{A}}[\mathbf{r}^{N}(t-\tau)]\mathbf{1}_{\mathrm{B}}[\mathbf{r}^{N}(t)]dt \, ,
\end{equation}
which is a time-scaled, joint probability of being in A at some time and being in B after time $\tau$.  This flux directly counts the number of times A transfers to B over a trajectory. If the lag-time $\tau$ is larger than the relaxation time within the A state, but much smaller than the typical waiting time for the transition, $q_{\mathrm{AB}}$ reports on the rate constant $k_{\mathrm{AB}}$ as $k_{\mathrm{AB}}=q_{\mathrm{AB}}/Y_{\mathrm{A}}$, independent of $\tau$ \cite{chandler1978statistical}.

In thermal equilibrium, probability fluxes are balanced, such that $q_{\mathrm{AB}}=q_{\mathrm{BA}}$ and the probability current $j_{\mathrm{AB}}\equiv q_{\mathrm{AB}}-q_{\mathrm{BA}}=0$. This implies that the ratio of forward and backward rates of interconversion $k_{\mathrm{AB}}$ and $k_{\mathrm{BA}}$ is strictly coupled to the relative stability of A and B, $k_{\mathrm{AB}}/k_{\mathrm{BA}}=Y_{\mathrm{B}}/Y_{\mathrm{A}}$. A configuration that is more energetically stable is necessarily less reactive at equilibrium, and thus the rate of error correction and annealing towards the globally stable state is reduced. This trade-off is the reason for the existence of an optimal zone for self-assembly in or near thermal equilibrium \cite{whitelam2015statistical}. Far-from-equilibrium conditions may be designed to stabilize otherwise unstable or exotic structures, but may not always break detailed balance, if the driving force does not couple to the relevant reactive mode. We describe next how to design far-from-equilibrium steady-states to tune rate constants and structural stabilities independently, and how to break detailed balance in any target coordinate.

\subsection*{Optimal design of function}
We use an iterative optimization algorithm to find the best choice of design parameters $\{D_{ij}\}$ and shear flow rate $f$ that maximizes fluxes $q_{\mathrm{AB}}$ or currents $j_{\mathrm{AB}}$ for a given choice of states A and B. We construct a cost function for optimization, $\langle\Omega\rangle$, that encodes the minimal change to a reference system to maximize either observable by minimizing the Kullback–Leibler divergence between the steady-states of the reference and the controlled system \cite{das2021variational}. For the overdamped dynamics we consider, this is given by $\langle\Omega\rangle=\left\langle \lambda O-\sum_{i=1}^{N}(\mathbf{u}_{i}-\mathbf{F}_{i})^{2}/4\gamma k_{\mathrm{B}}T\right\rangle $, where $\lambda$ is a biasing parameter, $O$ is a time-averaged static or dynamic order parameter, $q_{\mathrm{AB}}$ or $j_{\mathrm{AB}}$ for example, $\mathbf{u}_{i}$ is the total force that the $i$-th particle experiences, $\mathbf{F}_{i}$ is the WCA force from the interparticle potential, and $\langle \dots \rangle$ denote an average over the steady-state. 
The second term, deriving from the Kullback-Leibler divergence between trajectory ensembles with and without the added force $\mathbf{u}_i-\mathbf{F}_i$ \cite{das2019variational}, acts as a regularizer for the optimization in regions of design space where the first term is either zero or degenerate.  We optimize $\langle\Omega\rangle$ through stochastic gradient descent by computing its explicit gradients with respect to any design parameter $c$, from a generalized fluctuation-response relation
\begin{equation}\label{eq:grad}
    \frac{\partial\langle\Omega\rangle}{\partial c}=\left\langle \frac{\partial\Omega}{\partial\mathbf{u}}.\frac{\partial\mathbf{u}}{\partial c}\right\rangle +\int_{0}^{\infty}\left\langle \delta\Omega(t)\delta\left( \frac{\partial\dot{\ln}P[X]}{\partial \mathbf{u}}.\frac{\partial\mathbf{u}}{\partial c}\right) (0)\right\rangle  dt
\end{equation}
where $P[X]$ is the Onsager-Machlup probability of trajectory $X$\cite{Onsager_Machlup,warren2012malliavin}. We choose initial values for the design parameters such that $\langle O\rangle$ is nonzero, simulate a steady-state molecular dynamics trajectory to converge the terms in Eq.~\ref{eq:grad}, change the design parameters by taking a gradient descent step, and keep iterating till we converge to a locally optimal design. In practice, we find that as the gradients are stochastic, noise enhances the exploration of parameter space and helps anneal rapidly into the global optimum \cite{das2021variational}. We nevertheless test the uniqueness of each solution basin by starting the optimization from different points in parameter space, such as from high-yield designs for $\mathrm{A}$ and $\mathrm{B}$, or from a nonspecific attraction that results in moderate yields of both. During optimization, we constrain the magnitude of the bond energies to stay less than $10k_{\mathrm{B}}T$ so that there are no long-lived kinetic traps that prevent the relaxation of the system into a steady-state \cite{hormoz2011design}. In case of a breakdown in ergodicity, the steady-state gradient expression in Eq. \ref{eq:grad} would not be accurate, instead the gradients of probability of finite duration trajectories would need to be used \cite{das2022direct}.
Further details about the variational algorithm and a pseudocode are in the Supporting Information (SI).

\subsection*{Symmetric and asymmetric strategies for maximal flux}
We first apply this algorithm to the interconversion between polytetrahedral (A: $\mathrm{C}_{\mathrm{2v}}$) and octahedral (B: $\mathrm{O}_{\mathrm{h}}$) configurations in a sheared steady-state with $N=6$ particles, for a fixed choice of lag-time $\tau=0.025t^{*}$, slightly longer than typical transition timescales. The optimal design that maximizes yields $Y_{\mathrm{A}}$ and $Y_{\mathrm{B}}$ are known to be the \textit{Maximal Alphabet} $\mathbf{A}$ and $\mathbf{B}$ for the $D_{ij}$ matrix, shown schematically in Fig. \ref{fig:fig1}b, in which a uniform attractive energy stabilizes each contact pair \cite{das2021variational,hormoz2011design}. \textit{Alphabet} matrices possess a degeneracy to permutations of particle indices, a symmetry that the variational optimization spontaneously breaks \cite{das2021variational}. In order to interpret physically meaningful distinctions between alphabets and to interpolate between them, we have permuted them to be the least different (see SI for details). 

Optimization for maximal $q_{\mathrm{AB}}$ yields $f/f^{*}=50$, the highest allowed value during optimization, and two distinct $D_{ij}$ alphabet solutions, denoted $\mathbf{S}$ and $\mathbf{AS}$. Maximizing $q_{\mathrm{BA}}$ however yields only the alphabets $\mathbf{S}$. Figure \ref{fig:fig1}d shows a slice through the design space in the direction of varying $f$ but keeping $D_{ij}$ fixed. Far-from-equilibrium conditions amplify the fluxes by increasing the rates of interconversion in both directions. At $f=0$, detailed balance constraints $q_{\mathrm{AB}}=q_{\mathrm{BA}}$ as seen by the trade-offs in $(Y_{\mathrm{A}},k_{\mathrm{AB}})$ and $(Y_{\mathrm{B}},k_{\mathrm{BA}})$. The $\mathbf{S}$ matrix maximizes both $q_{\mathrm{AB}}$ and $q_{\mathrm{BA}}$ symmetrically, such that probability flows predominately between states A and B even far-from-equilibrium. This is accomplished with alphabet $\mathbf{S}$ being intermediate in bond-strength to the alphabets $\mathbf{A}$ and $\mathbf{B}$. $\mathbf{S}$ preserves the common bonds in $\mathbf{A}$ and $\mathbf{B}$ and codes for weak bond energies for the unique bond in each structure. 

With the other solution $\mathbf{AS}$, shear flow amplifies the flux in the forward direction more than the reverse, breaking detailed balance. Figure \ref{fig:fig1}e shows how this is achieved by starting with a $Y_{\mathrm{A}}$ higher than the previous case and thus a smaller $k_{\mathrm{AB}}$, at equilibrium. At high shear, the $k_{\mathrm{AB}}$ is amplified to similar absolute values as with $\mathbf{S}$, without a proportional reduction in $Y_{\mathrm{A}}$. 
The solution $\mathbf{AS}$ codes for a dynamical bonding arrangement different from $\mathbf{A}$ or $\mathbf{B}$. This class of asymmetric high-flux solutions is generally found as a locally optimal design for other pairs of nanoclusters as well. We note that the alphabet in question is not asymmetric, as it is generated from pairwise conservative interactions, but rather that the alphabet causes $q_{\mathrm{AB}}$ and $q_{\mathrm{BA}}$ to be asymmetrically amplified, breaking detailed balance. 

In both cases, the fluxes reach a maximum with $f$ before turning over and decaying at large shear rates that are nevertheless experimentally achievable in microfluidic devices \cite{dou2017review}. This turnover arises from a competition between the enhancement of diffusive exploration of the internal degrees of freedom till moderate flow rates, and the destabilization of bonds at high flow rates. This turnover is similar in origin to a turnover previously reported in shear-induced crystallization rates of hard spheres at high supersaturation, where the cohesive free energy is much larger than $k_{\mathrm{B}}T$ \cite{richard2015role}.

\begin{figure*}[t]
\centering
\includegraphics[trim={3cm 0 0 0},width=17.8cm]{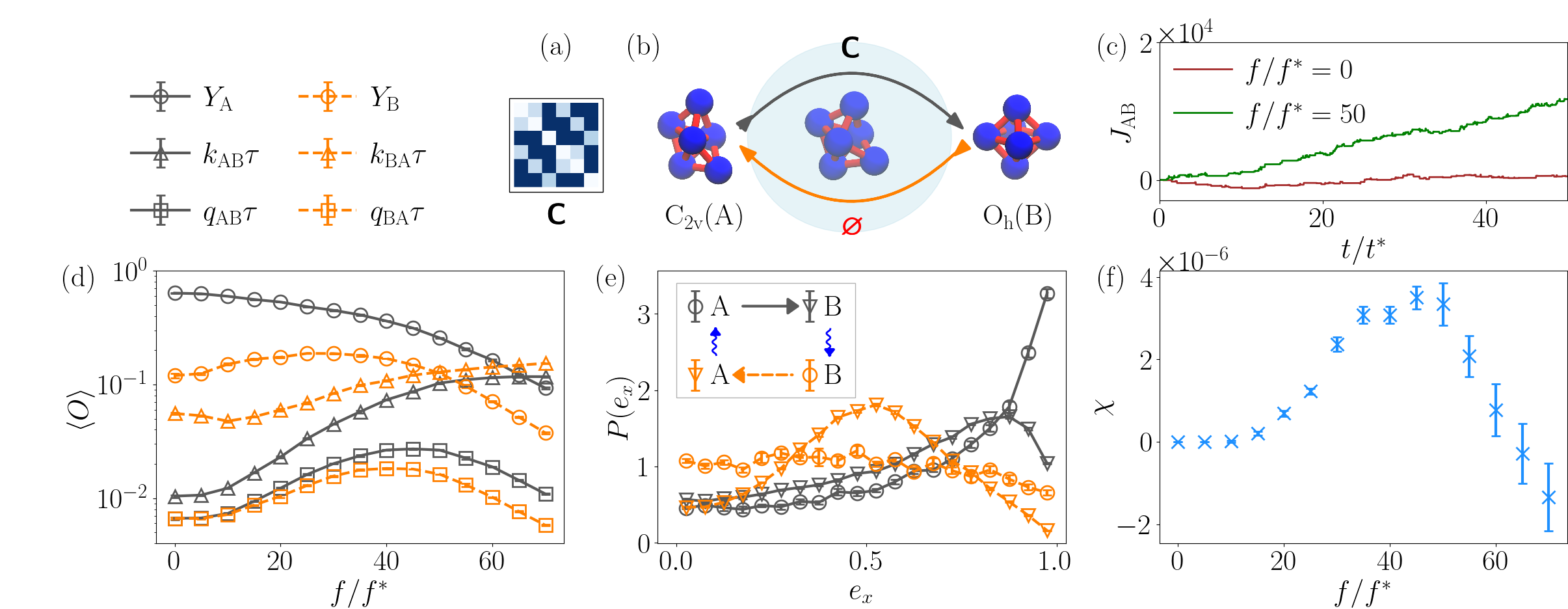}
\caption{Optimal breakdown of detailed balance induced by shear flow. (a) Optimized alphabet for high current. (b) Schematic of the isomerization intermediate. (c) Reactive displacement trajectories $J_{\mathrm{AB}}(t)$. (d) Average values of observable $\langle O\rangle\in\{Y,k\tau,q\tau\}$ with respect to shear, keeping the alphabet fixed at $\mathbf{C}$. (e) Probability distribution of the orientation of the nanocluster, $e_{x}$, while in steady-state in $\mathrm{A}$ (black circles), immediately after an $\mathrm{A}\to\mathrm{B}$ reaction (black triangles), steady-state in $\mathrm{B}$ (orange circles), and immediately after a $\mathrm{B}\to\mathrm{A}$ reaction (orange triangles). Blue arrows in inset denote orientational relaxation.
(f) Difference of covariance of shear work $W$ and the time spent in 11-bonded state $\tau_{\mathrm{I}}$, during forward and backward transitions. 
}
\label{fig:fig2}
\end{figure*}

\subsection*{Strategy for maximal probability current}
For the same polytetrahedron (A) and octahedron (B) clusters, we have also found the optimal $D_{ij}$ matrices and shear flow rates $f$ that maximize the probability currents $j_{\mathrm{AB}}$ and $j_{\mathrm{BA}}$. For $j_{\mathrm{AB}}$, we find the optimal solution to be $f/f^{*}$ close to $50$ and alphabet $\mathbf{C}$, shown in Fig. \ref{fig:fig2}a. 
%
%
%
For $j_{\mathrm{BA}}$ optimization, we do not find any design where $j_{\mathrm{BA}}$ is positive and the optimization algorithm always finds the trivial maximum with low $D_{ij}$ where $Y_{\mathrm{A}}=Y_{\mathrm{B}}=0$. A slice through the parameter space in the $f$ direction, keeping alphabet $\mathbf{C}$ fixed, is shown in Fig. \ref{fig:fig2}d. The algorithm has found a design to break detailed balance maximally in this pre-specified direction, by decoupling constraints between $(Y_{\mathrm{A}},k_{\mathrm{AB}})$ and $(Y_{\mathrm{B}},k_{\mathrm{BA}})$. When we compare these observables to those with the symmetric high-flux alphabet $\mathbf{S}$ in Fig. \ref{fig:fig1}d, we find that the high current with $\mathbf{C}$ arises due to $Y_{\mathrm{B}}$ having a quicker turnover than before. An equivalent explanation is $k_{\mathrm{BA}}$ does not increase as expected from detailed balance considerations. As a result, structural constraints have been decoupled optimally from rate constraints in nonequilibrium. The gap between $Y_{\mathrm{A}}$ and $Y_{\mathrm{B}}$ in log-scale, exactly balanced at equilibrium by the gap between $k_{\mathrm{AB}}$ and $k_{\mathrm{BA}}$, has been designed to no longer be so at the optimal shear.

Broken detailed balance results in a \textit{reactive displacement} over time $J_{\mathrm{AB}}(t)=\int_{\tau}^{t+\tau}\left[ \mathbf{1}_{\mathrm{A}}(t^{'}-\tau)\mathbf{1}_{\mathrm{B}}(t^{'})-\mathbf{1}_{\mathrm{B}}(t^{'}-\tau)\mathbf{1}_{\mathrm{A}}(t^{'})\right] dt^{'}$,
where the full arguments of the indicators have been suppressed for brevity. The expectation of this observable counts the difference between the expected number of transitions happening in the forward and backward directions. In Fig. \ref{fig:fig2}c we have shown a typical trajectory of the reactive displacement with time, at zero and optimal shear rate, for the alphabet $\mathbf{C}$. The approximately uniform and rapid growth of $J_{\mathrm{AB}}$ with time, and the comparable magnitudes of fluctuations in the two curves, shows how the shear is preferentially amplifying the reactive flux in the forward direction and that this happens through a higher number of small jumps rather than through large rare events. This suggests that the reactive \textit{cycle} involved in breaking detailed balance in the $\mathrm{A}\to\mathrm{B}$ coordinate consists of all edges with comparable reaction rates and comparable changes in configuration space.

Shear flow is a nonspecific mechanical force acting on all particles, yet it is able to drive a reactive cycle in the internal coordinates of the nanocluster. This is achieved through autonomous relaxations in the orientation of the nanoclusters. We demonstrate this by plotting probability distributions of the x-component of the normalized principal eigenvector of the gyration tensor of the nanocluster, $e_{x}$, in Fig. \ref{fig:fig2}e. The $x$ component of the gyration tensor $e_{x}$ measures the orientational alignment of the cluster in the flow direction. Aspherical clusters like $\mathrm{A}$ have larger typical value of $e_{x}$ than approximately spherical clusters like $\mathrm{B}$, which does not align in the flow. We find that immediately after reactions in both directions, orientations of the clusters are significantly different from their steady-state values. Orientational motion aided by shear flow and conformational transitions thus form a reactive cycle with a nonzero current.

The distinction between the forward and backward transition paths  with alphabet $\mathbf{C}$ in the presence of shear flow can be quantified mechanistically.  Both clusters have 12 bonds, but to transition between the two states requires breaking a bond on the concave face of the $\mathrm{C}_{\mathrm{2v}}$ structure, passing through an 11-bonded intermediate as shown in Fig. \ref{fig:fig2}b, and then rebonding of two different particles on that face. The optimized alphabet $\mathbf{C}$ stabilizes this motion by weakening the bond in $\mathrm{A}$ located directly opposite to its concave face. We have computed the effect of shear flow on the transition paths through the covariance between the time spent in the 11-bonded intermediate, $\tau_{\mathrm{I}}$, and the work done by shear flow, $W=\sum_{i=1}^{N}\int_{t_{1}}^{t_{2}}\mathbf{f}_{i}^{\mathrm{S}}(\mathbf{r}_{i}).\dot{\mathbf{r}_{i}}dt$, which is integrated between $t_{1}$ and $t_{2}$, the start and end of forward and backward transition paths. The difference in covariance 
$
\chi  = \langle \delta \tau_{\mathrm{I}} \delta W \rangle_\mathrm{AB}- \langle \delta \tau_{\mathrm{I}} \delta W \rangle_\mathrm{BA}
$
between forward and reverse reaction paths, is plotted in Fig. \ref{fig:fig2}f, with respect to increasing shear. The difference in the covariance increases initially and then peaks near the optimal shear flow for high current. This implies that even at the level of individual trajectories, spending time in the 11-bonded intermediate state is correlated with the effect of shear more in the trajectory ensemble of forward reactions than the reverse ones. Past the turnover, the difference becomes smaller and then negative, with fluctuations that grow with further increasing shear, suggesting participation of competing sheared reaction pathways involving more broken bonds.  



\begin{figure}[t]
\centering
\includegraphics[width=8.3cm]{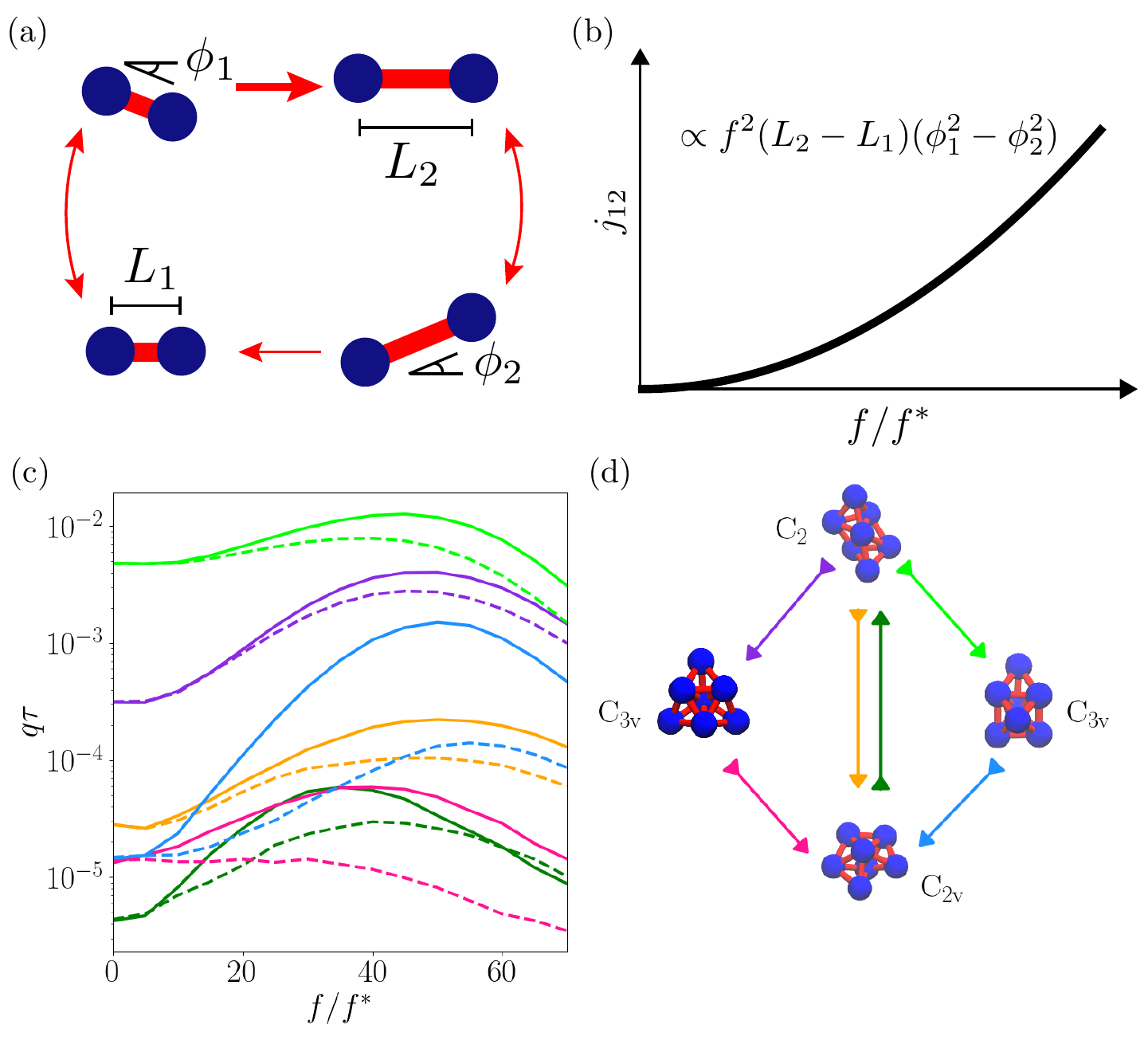}
\caption{Design for maximal probability current. (Top row) Minimal model of shear-induced reactive cycle for a two-dimensional colloidal dimer. (a) Reactive cycle comprising of shear-dependent reactions followed by shear-dependent orientational relaxations. (b) Quadratic response of the probability current to an increasing shear flow. (c) Profile of reactive currents in 7-particle clusters with changing shear, keeping the optimized alphabets fixed. Solid (dashed) lines represent forward (backward) reactive flux. Colors correspond to the specific edges in the reaction network in (d).}
\label{fig:fig3}
\end{figure}

Detailed studies of the local stability of these strategies in design space (SI) show that detailed balance is preserved all along linear interpolations in parameter space between $\mathbf{A}\to\mathbf{S}\to\mathbf{B}$ even in a high shear flow, with the highest flux occurring at $\mathbf{S}$ as found by the optimization algorithm. 
Employing similar interpolations between $\mathbf{S}\to\mathbf{AS}\to\mathbf{C}$, we find that $q_{\mathrm{AB}}$ increases monotonically between $\mathbf{C}\to\mathbf{AS}$, which explains why $\mathbf{AS}$ can be discovered by gradient descent when starting from $\mathbf{C}$.
The close connection between solutions $\mathbf{C}$ and $\mathbf{AS}$ suggests that the existence of a nontrivial high current locally optimal solution in any reaction direction depends on the existence of an asymmetric high flux solution with a positive current.

\begin{figure*}[t]
\centering
\includegraphics[trim={4cm 0 0 0},width=17cm]{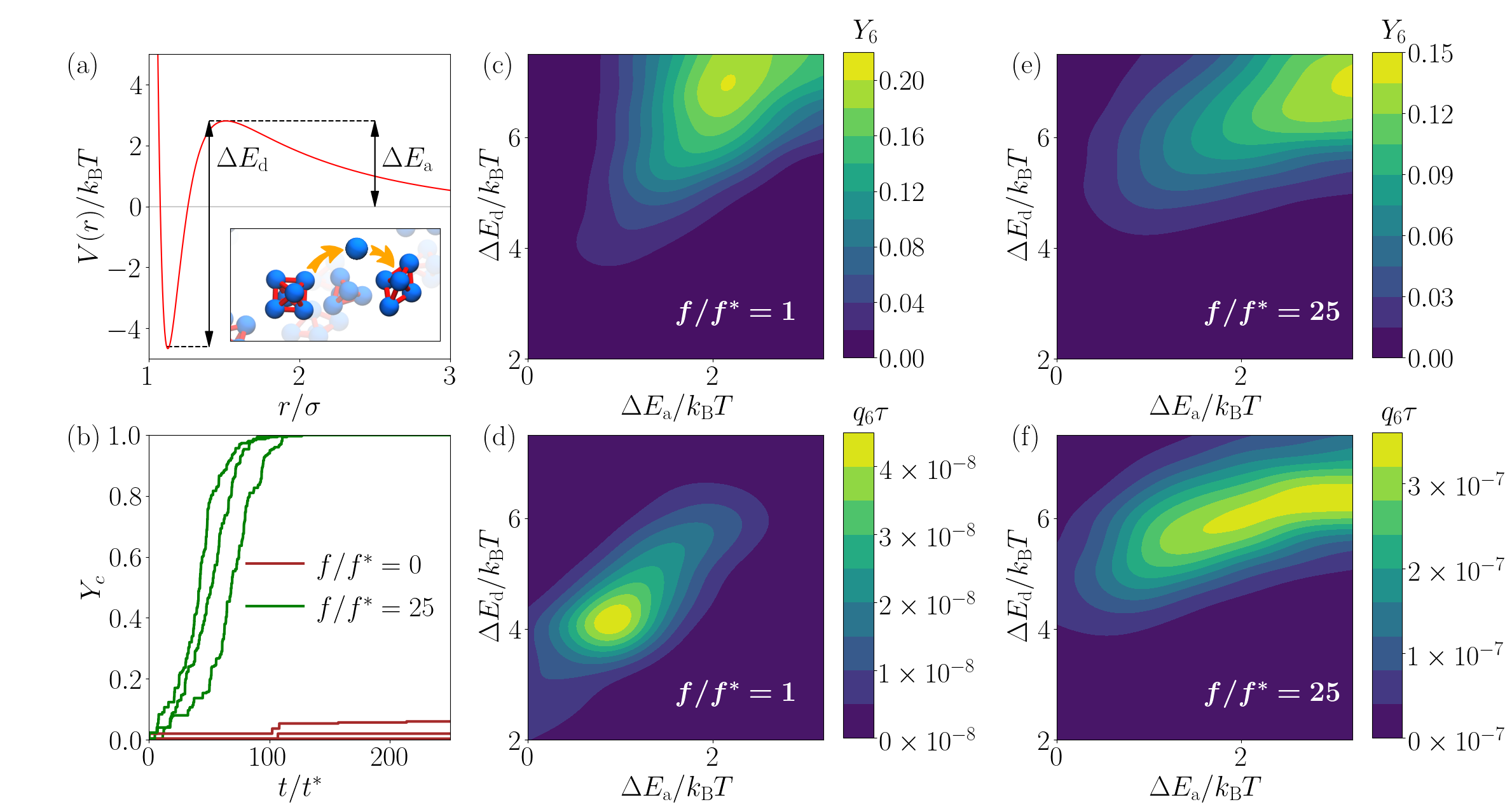}
\caption{Independent tuning of hopping flux and microphase structure by shear flow. (a) Pairwise interaction potential. (Inset) Schematic illustration of hopping in a microphase. (Center and right columns) Color maps of (c) $Y_{6}$ at $f/f^{*}=1$, (d) $q_{6}\tau$ at $f/f^{*}=1$, (e) $Y_{6}$ at $f/f^{*}=25$ and (f) $q_{6}\tau$ at $f/f^{*}=25$, with respect to changing assembly and disassembly barriers. Corresponding color scales are indicated next to each color map. (b) Trajectories of the yield of particles of color $c$ as a function of time, with different noise realizations.}
\label{fig:fig4}
\end{figure*}

\subsection*{Generality of nonequilibrium strategies}
We can better understand the coupling between shear-driven orientation and the conformational change through a minimal model of a colloidal dimer in shear flow $\mathbf{f}_{i}^{\mathbf{S}}=fz_{i}\mathbf{x}_{i}$ constrained to the $x-z$ plane for simplicity. We assume that the dimer isomerizes between two states $\alpha_{1}$ and $\alpha_{2}$, with center-to-center distances $L_{1}$ and $L_{2}$ respectively, and equilibrium rates  $k_{12}=k_0$ and $k_{21}=k_0 e^{\beta \Delta V}$, where $\Delta V$ is the potential energy difference between $\alpha_{2}$ and $\alpha_{1}$. This is schematically illustrated in Fig. \ref{fig:fig3}a. We assume $L_{2}=L_{1} + \Delta L$, such that a shear flow may stretch the dimer and increase the rate of isomerizing from $\alpha_{1}$ to $\alpha_{2}$. We assume the reactions occur at fixed orientation, and in between reactions, the orientation of the dimer relaxes in the shear flow.
The orientation is quantified by the angle $\theta\in(0,2\pi)$ where $\cos\theta\equiv\hat{\mathbf{e}} \cdot \hat{\mathbf{x}}$ with $\hat{\mathbf{e}}$ being the dimer axis. 
%
%
The torque due to the shear flow depends on the state $\alpha_{j\in(1,2)}$ and is given by $\Gamma=fL_{j}^{2}[1-\cos(2\theta)]/4$, which together with an angular diffusion constant $D_{\theta}$, relaxes the distribution $P(\theta)$. In the limit of small $fL_{j}^{2}/D_{\theta}$, the distribution $P_{j}(\theta)$ for each state $\alpha_{j\in(1,2)}$ can be computed as a perturbation over free diffusion, $P_{j}(\theta)=\{1+[8D_{\theta}p\sin2\theta-p^{2}(\cos4\theta-4\cos2\theta) ]/16D_{\theta}^{2}\}N_{j}/2\pi$, where $p=fL_{j}^{2}/4$ and $N_{j}$ is a normalization factor \cite{gardiner1985handbook}. Longer clusters thus align more with shear flow, similar to Fig. \ref{fig:fig2}e. We further assume that the yields of $\alpha_{1}$ and $\alpha_{2}$ do not change from equilibrium, such that $N_{1}/N_{2}=e^{\beta \Delta V}$, consistent with our explicit model findings.

Given an orientation, the stretching force is $\Pi=fL_{j}\sin(2\theta)/4$ and upto linear order in $\Delta L$, the rates of extension and compression are affected as $k_{12}=k_0 e^{\beta \Pi\Delta L\cos\phi_{1}}$ and $k_{21}=k_0e^{\beta (\Delta V+\Pi\Delta L\cos\phi_{2})}$ respectively \cite{kuznets2021dissipation}, where $\phi_{j\in(1,2)}$ is the angle between the reaction coordinate and the stretching mode. We can now compute the forward and backward fluxes as probability times reaction rate integrated over all $\theta$. The steady-state reactive current from $\alpha_{1}$ to $\alpha_{2}$ becomes
\begin{equation}
j_{12}=\frac{k_0 L^{3}}{64D_{\theta}^{2} [1+e^{-\beta \Delta V}]}\Delta L(\cos\phi_{1}-\cos\phi_{2})f^{2}
\end{equation}
in the limit of small $\Delta L/L$ where $L=(L_{1}+L_{2})/2$. The quadratic response of the current in the limit of small $\phi_{j}$ is illustrated in Fig. \ref{fig:fig3}b.
Hence shear flow is able to produce a reactive current by coupling internal conformation with global orientation in a four-stroke cycle, with the sign of the current depending only on the geometry of the reaction coordinates. While simple, this model explains the observed features in the self-assembly of nanoclusters, such as why there exist no solutions for positive current in some reaction directions. Based on our minimal model, bidirectional currents are only possible when there exist multiple reaction mechanisms, \textit{i.e.}, different sets of $(\phi_{1},\phi_{2})$. Further the dimer model predicts that the magnitude of the current $j_{12}$ increases with increasing $\Delta V$, \textit{i.e.}, with a more stable $\alpha_{1}$, when the timescale arising from $k_0$ is held constant. This rationalizes the observation that the alphabet $\mathbf{C}$ in Fig. \ref{fig:fig2}d that drives current $j_{AB}$ has a higher $Y_{A}/Y_{B}$ at equilibrium, compared to alphabet $\mathbf{S}$. 

A high reactive flux simultaneously in both directions is achieved through having a low free-energy barrier such that the reaction timescale is faster than shear-induced rotations, with the fluxes being amplified symmetrically by a shear flow. A high reactive current is achieved through having a favorable reaction coordinate geometry and an intermediate barrier height, to slow down the reactive timescale and couple it to to rotational timescale. For nanocluster rearrangements where the free energy barrier and reaction coordinate geometry are not available, the variational algorithm can directly optimize for the reactive flux and current.
We note that this method of converting mechanical driving to chemical work through switchable geometry is reminiscent of the blood-clotting protein von Willebrand factor that uses the shear flow to unfold \cite{schneider2007shear}. The shear-induced enhancement of reactive flux and current in our theory is contingent only on two aspects, the alignment of aspherical clusters along a shear flow, and the modulation of reaction rates by dissipative driving. Both are general results with numerical and experimental verification in colloidal suspensions \cite{vermant2005flow,toyabe2007experimental,kuznets2021dissipation}, polymers \cite{wu2006shear,alexander2006shear} and biomolecules \cite{hill2006shear,liphardt2002equilibrium}.

The generality of the principle can be  tested by studying  probability fluxes and currents for interconversions between four rigid nanoclusters with $N=7$ DNA-labeled particles, as shown in Fig. \ref{fig:fig3}c) and d), with additional results in the SI. These are the four highest yield clusters at equilibrium for a nonspecific short-range pairwise attraction \cite{das2021variational,hormoz2011design}. We find that the nonequilibrium strategies discovered for interconverting the polytetrahedron and octahedron nanoclusters are also applicable more widely. 
When pairwise interaction matrices are set to the optimized alphabets, most reactive fluxes are amplified by a far-from-equilibrium shear flow due to a larger enhancement in rate constants compared to the decay in yields. 
The far-from-equilibrium optima exhibit qualitatively different dynamics than a near-equilibrium perturbative regime. For high probability flux in pairwise transitions, in most cases we find locally optimal symmetric alphabets that preserve detailed balance even far-from-equilibrium. The symmetric solutions are always locally optimal in both forward and reverse directions of the transitions. As before, these solutions result from restricting transitions between only these two tagged configurations, and preventing leakage to competing structures.  In many cases we find distinct locally optimal asymmetric alphabets that break detailed balance at large $f$. The alphabets for this strategy are typically only optimal for reactive flux in one reaction direction. When optimized for high probability current, we find many locally optimal alphabets that channel the shear flow into breaking detailed balance in a particular reaction direction, and thus maximizes the difference between forward and backward reactive flux.  These alphabets are distinct from, but closely related to, the asymmetric high-flux alphabets.  For any given reaction, optimal high-current solutions exist if and only if there are optimal asymmetric high-flux solutions that have a positive current.  


\subsection*{Particle hopping flux in a sheared microphase}
Extending these strategies to a macroscopic system of DNA-labeled colloids straightforwardly requires a prohibitive $\mathcal{O}(N^{2})$ scaling of the necessary alphabet size, as any colloid in a distinct nanocluster must not attract bonding partners from adjacent nanoclusters \cite{hormoz2011design,zeravcic2014size}. This attraction is a nanoscale manifestation of surface tension that tries to relax the system towards a bulk condensate. If however a long-range repulsion arising from a screened Coulomb interaction is combined with the short-range DNA-labeled attractive forces, large systems can self-assemble into a  \textit{microphase} at equilibrium, characterized by the spontaneous emergence of an intermediate length-scale at which nanoclusters are stabilized \cite{zhuang2016recent, hagan2021equilibrium}. Microphase separation or self-limited assembly is often used in biological systems to regulate the volumes of viral capsids, cells and membraneless organelle \cite{caspar1962physical,tanaka2008atomic,courchaine2016droplet}. 

Self-limited clusters in biological systems are not static structures but highly fluxional. Nonequilibrium surface reactions are used by cells to stabilize microphase droplets \cite{zwicker2015suppression,soding2020mechanisms,kirschbaum2021controlling}, and components of self-limited clusters continuously exchange with the surrounding solute pool in a dynamic steady-state \cite{brackley2017ephemeral}. Here we focus on how to tune the dynamics of particle exchange between nanoclusters given we have a microphase separated state. By studying the optimal design of a particle hopping flux between multiple nanoclusters, we demonstrate that a sheared nonequilibrium steady-state can break equilibrium constraints relating stability and flux.

We start with a microphase separated state at equilibrium, self-assembled from DNA-labeled colloids  having a pairwise repulsive Yukawa potential $V(r)=\epsilon\exp(-\kappa r)/r$ in addition to the previous short-ranged interactions. We are interested in the yield and reactive flux of compact nanoclusters consisting of 6 colloids and with at least 9 bonds, regardless of its shape. Such nanoclusters are denoted as state A. The cutoff on the number of bonds ensures we are accurately counting reactions involving hopping of particles between clusters instead of complete dissociation of a cluster. We calculate the average yield $Y_{\mathrm{6}}$ at any time $t$, by checking whether each particle is in a cluster of type A, as $Y_{\mathrm{6}}=\int_{0}^{t_{f}}\sum_{i=1}^{N}\mathbf{1}_{i\in\mathrm{A}}(t)dt/Nt_{f}$. For counting particle hops we take a lag-time $\tau=0.025t^{*}$ slightly larger than the timescale of typical adsorption and desorption transitions of colloids from nanoclusters. With this, our hopping flux between 6-colloid clusters becomes
\begin{equation}
    q_{6}=\frac{1}{N\tau(t_{f}-\tau)}\int_{\tau}^{t_{f}}\sum_{i=1}^{N}\mathbf{1}_{i\in\mathrm{A}^{'}}(t-\tau)\mathbf{1}_{i\in\mathrm{A}^{''}}(t)dt
\end{equation}
where $\mathrm{A}^{'}$ and $\mathrm{A}^{''}$ refer to distinct clusters of type A with the only common particle being that with index $i$. This condition ensures that only the exchange of a single particle at a time, \textit{i.e.} a \textit{hop}, is being counted, as illustrated in Fig. \ref{fig:fig4}a (\textit{inset}).

As we are not recording the shapes of clusters so long that they are compact, we do not use an alphabet in our design space, but rather a nonspecific value of the attraction $D$ that competes against the repulsion energy scale $\epsilon$ over a lengthscale of $\kappa^{-1}$. We constrain the total pairwise force and potential to go to zero at a cutoff distance of $5\sigma$ with a shifted forces approximation \cite{toxvaerd2011communication}. This constraint leaves the potential energy function with two relevant effective features, the energy barrier to bond formation or the assembly barrier $\Delta E_{\mathrm{a}}$, and the energy barrier to bond breaking or the disassembly barrier $\Delta E_{\mathrm{d}}$, shown in Fig. \ref{fig:fig4}a. We study the effect of far-from-equilibrium conditions on $Y_{6}$ and $q_{6}$ by variationally optimizing $D$, $\epsilon$ and $\kappa$, keeping the shear flow rate fixed at $f/f^{*}=1$ and 25. We also compute the yield and reactive flux landscape by directly sweeping over the design parameters such that we can elucidate their dependence on $\Delta E_{\mathrm{a}}$ and $\Delta E_{\mathrm{d}}$. 

\subsection*{Breaking equilibrium constraints on hopping flux}
At equilibrium, large values of both barriers $\Delta E_\mathrm{d}$ and $\Delta E_\mathrm{a}$ stabilizes the microphase, where their difference $\Delta E_\mathrm{d}-\Delta E_\mathrm{a}$ is the energy gain on forming a new bond starting from a dispersed state. The reactive flux for the bond formation and breakage, however, is dependent on both of these energy barriers, slowing exponentially as they increase. As can be anticipated from detailed balance, near equilibrium changing $\Delta E_\mathrm{d}$ or $\Delta E_\mathrm{a}$ to optimize yield degrades the reactive flux and vice-versa. This trade-off is shown near thermal equilibrium, $f/f^*=1$, in Fig. \ref{fig:fig4}c and \ref{fig:fig4}d in maps of $Y_{6}$ and $q_{6}$, for a system of $N=90$ colloids at a packing fraction $\phi=0.02$, where the yield and reactive flux both exhibit narrow peaks at distinct parameters. 

Far-from-equilibrium, however, the constraint between yield and reactive flux is decoupled, and both have broad regimes of near optimal behavior that overlap one another. Shear flow thus allows an optimal regime in parameter space having high yield of a microphase as well as a high reactive flux of particles hopping between nanoclusters.  This occurs by expanding the regime of stability of the microphase over a broader $\Delta E_\mathrm{a}$ range than near equilibrium (see Fig. \ref{fig:fig4}e and \ref{fig:fig4}f) through stabilizing nonrigid nanoclusters \cite{das2021variational}. These nonrigid clusters are compact with $\geq 9$ bonds in total, but have dangling bonds that are energetically more susceptible to break, contributing to a high hopping flux.   We find that though far-from-equilibrium shear flow reduces the yield of the nanoclusters by 30\% from the near-equilibrium regime, the highest reactive flux is amplified by more than 600\%. This points at a large amplification in the underlying rate constants by the breakdown of equilibrium constraints. We also show in the SI that the enhancement of hopping flux is explained by the alignment of nonrigid clusters in the shear flow and the enhancement of diffusion constant in the flow direction. These findings are in qualitative agreement with the shear-induced flux enhancement in our dimer model. In all cases the variational algorithm optimizes the parameters accurately to arrive close to the optima given it is started from a point in parameter space where either observable is measurably non-zero (SI).

We have demonstrated an application of the increased hopping flux between nanoclusters by an example of an autocatalytic switching reaction. We assume that colloids in the nanoclusters are energetically identical but each carry a color, $c$ or $nc$, similar to phosphorylation tags on biomolecules or competing conformations in proteins \cite{brackley2017ephemeral,michaels2020dynamics}. At $t=0$ in a steady-state only one particle is initialized as $c$ with the rest being $nc$. Every time a $c$-colored particle is in a cluster of type A, it instantly converts every other particle in the cluster to $c$, similar to models for cooperative switching in biomolecules \cite{sourjik2004functional}. The yield $Y_{c}$ of particles of color $c$ is expected to increase fastest when colloids from type A clusters hop fastest directly to other type A clusters in a steady-state. Keeping $D$, $\epsilon$ and $\kappa$ fixed at the optimal value for the highest microphase yield $Y_{6}$ at $f/f^{*}=1$, we show example trajectories of $Y_{c}$ for $f/f^{*}=0$ and $25$ with a system size $N=300$, in Fig. \ref{fig:fig4}b. We find that the far-from-equilibrium system indeed shows a higher rate of increase of $Y_{c}$ over the equilibrium scenario. At this optimal design the mechanical and nonspecific shear driving has been optimally channeled into amplifying reactive flux in the hopping reaction coordinate. 

We note that the reactive fluxes we discussed are autonomously driven within the sheared steady-state assembly. The design principles we found rely on coupling an external nonspecific driving force to an internal reaction coordinate through an autonomously self-organizing degree of freedom, the geometry. The accompanying algorithm can numerically discover such design rules.
Our insights on how such motifs can break equilibrium constraints in specific reaction coordinates can help design molecular machines that efficiently transduce energy through internal rearrangements \cite{seifert2012stochastic}. This work also opens a promising window towards designing dynamical phases of functional materials, and shows how to tune fluxionality without compromising structure, by going to nonperturbative far-from-equilibrium regimes \cite{whitelam2012self}.




{\bf Materials and Methods} A pseudocode of our optimization algorithm, a detailed description of the numerical simulations, learning curves, and optimal alphabets are available in the SI.

{\bf Acknowledgements} This work has been supported by NSF Grant CHE-1954580. AD was also supported by the Philomathia Graduate Student Fellowship from the Kavli Energy Nanoscience Institute at UC Berkeley.

{\bf Data availability} A source code for the optimization algorithm and all data that reproduce the findings of this study are openly available on Zenodo at \href {https://doi.org/10.5281/zenodo.8031998}{https://doi.org/10.5281/zenodo.8031998}.\cite{das_noneq_functional_design_zenodo}


\begin{thebibliography}{60}%
\makeatletter
\providecommand \@ifxundefined [1]{%
 \@ifx{#1\undefined}
}%
\providecommand \@ifnum [1]{%
 \ifnum #1\expandafter \@firstoftwo
 \else \expandafter \@secondoftwo
 \fi
}%
\providecommand \@ifx [1]{%
 \ifx #1\expandafter \@firstoftwo
 \else \expandafter \@secondoftwo
 \fi
}%
\providecommand \natexlab [1]{#1}%
\providecommand \enquote  [1]{``#1''}%
\providecommand \bibnamefont  [1]{#1}%
\providecommand \bibfnamefont [1]{#1}%
\providecommand \citenamefont [1]{#1}%
\providecommand \href@noop [0]{\@secondoftwo}%
\providecommand \href [0]{\begingroup \@sanitize@url \@href}%
\providecommand \@href[1]{\@@startlink{#1}\@@href}%
\providecommand \@@href[1]{\endgroup#1\@@endlink}%
\providecommand \@sanitize@url [0]{\catcode `\\12\catcode `\$12\catcode
  `\&12\catcode `\#12\catcode `\^12\catcode `\_12\catcode `\%12\relax}%
\providecommand \@@startlink[1]{}%
\providecommand \@@endlink[0]{}%
\providecommand \url  [0]{\begingroup\@sanitize@url \@url }%
\providecommand \@url [1]{\endgroup\@href {#1}{\urlprefix }}%
\providecommand \urlprefix  [0]{URL }%
\providecommand \Eprint [0]{\href }%
\providecommand \doibase [0]{http://dx.doi.org/}%
\providecommand \selectlanguage [0]{\@gobble}%
\providecommand \bibinfo  [0]{\@secondoftwo}%
\providecommand \bibfield  [0]{\@secondoftwo}%
\providecommand \translation [1]{[#1]}%
\providecommand \BibitemOpen [0]{}%
\providecommand \bibitemStop [0]{}%
\providecommand \bibitemNoStop [0]{.\EOS\space}%
\providecommand \EOS [0]{\spacefactor3000\relax}%
\providecommand \BibitemShut  [1]{\csname bibitem#1\endcsname}%
\let\auto@bib@innerbib\@empty
\bibitem [{\citenamefont {Whitesides}\ and\ \citenamefont
  {Grzybowski}(2002)}]{whitesides2002self}%
  \BibitemOpen
  \bibfield  {author} {\bibinfo {author} {\bibfnamefont {G.~M.}\ \bibnamefont
  {Whitesides}}\ and\ \bibinfo {author} {\bibfnamefont {B.}~\bibnamefont
  {Grzybowski}},\ }\href@noop {} {\bibfield  {journal} {\bibinfo  {journal}
  {Science}\ }\textbf {\bibinfo {volume} {295}},\ \bibinfo {pages} {2418}
  (\bibinfo {year} {2002})}\BibitemShut {NoStop}%
\bibitem [{\citenamefont {Jacobs}\ \emph {et~al.}(2015)\citenamefont {Jacobs},
  \citenamefont {Reinhardt},\ and\ \citenamefont
  {Frenkel}}]{jacobs2015rational}%
  \BibitemOpen
  \bibfield  {author} {\bibinfo {author} {\bibfnamefont {W.~M.}\ \bibnamefont
  {Jacobs}}, \bibinfo {author} {\bibfnamefont {A.}~\bibnamefont {Reinhardt}}, \
  and\ \bibinfo {author} {\bibfnamefont {D.}~\bibnamefont {Frenkel}},\
  }\href@noop {} {\bibfield  {journal} {\bibinfo  {journal} {Proceedings of the
  National Academy of Sciences}\ }\textbf {\bibinfo {volume} {112}},\ \bibinfo
  {pages} {6313} (\bibinfo {year} {2015})}\BibitemShut {NoStop}%
\bibitem [{\citenamefont {Fullerton}\ and\ \citenamefont
  {Jack}(2016)}]{fullerton2016optimising}%
  \BibitemOpen
  \bibfield  {author} {\bibinfo {author} {\bibfnamefont {C.~J.}\ \bibnamefont
  {Fullerton}}\ and\ \bibinfo {author} {\bibfnamefont {R.~L.}\ \bibnamefont
  {Jack}},\ }\href@noop {} {\bibfield  {journal} {\bibinfo  {journal} {The
  Journal of Chemical Physics}\ }\textbf {\bibinfo {volume} {145}},\ \bibinfo
  {pages} {244505} (\bibinfo {year} {2016})}\BibitemShut {NoStop}%
\bibitem [{\citenamefont {Coropceanu}\ \emph {et~al.}(2022)\citenamefont
  {Coropceanu}, \citenamefont {Janke}, \citenamefont {Portner}, \citenamefont
  {Haubold}, \citenamefont {Nguyen}, \citenamefont {Das}, \citenamefont
  {Tanner}, \citenamefont {Utterback}, \citenamefont {Teitelbaum},
  \citenamefont {Hudson} \emph {et~al.}}]{coropceanu2022self}%
  \BibitemOpen
  \bibfield  {author} {\bibinfo {author} {\bibfnamefont {I.}~\bibnamefont
  {Coropceanu}}, \bibinfo {author} {\bibfnamefont {E.~M.}\ \bibnamefont
  {Janke}}, \bibinfo {author} {\bibfnamefont {J.}~\bibnamefont {Portner}},
  \bibinfo {author} {\bibfnamefont {D.}~\bibnamefont {Haubold}}, \bibinfo
  {author} {\bibfnamefont {T.~D.}\ \bibnamefont {Nguyen}}, \bibinfo {author}
  {\bibfnamefont {A.}~\bibnamefont {Das}}, \bibinfo {author} {\bibfnamefont
  {C.~P.}\ \bibnamefont {Tanner}}, \bibinfo {author} {\bibfnamefont {J.~K.}\
  \bibnamefont {Utterback}}, \bibinfo {author} {\bibfnamefont {S.~W.}\
  \bibnamefont {Teitelbaum}}, \bibinfo {author} {\bibfnamefont {{\c{}}.~M.~H.}\
  \bibnamefont {Hudson}},  \emph {et~al.},\ }\href@noop {} {\bibfield
  {journal} {\bibinfo  {journal} {Science}\ }\textbf {\bibinfo {volume}
  {375}},\ \bibinfo {pages} {1422} (\bibinfo {year} {2022})}\BibitemShut
  {NoStop}%
\bibitem [{\citenamefont {Whitelam}\ and\ \citenamefont
  {Jack}(2015)}]{whitelam2015statistical}%
  \BibitemOpen
  \bibfield  {author} {\bibinfo {author} {\bibfnamefont {S.}~\bibnamefont
  {Whitelam}}\ and\ \bibinfo {author} {\bibfnamefont {R.~L.}\ \bibnamefont
  {Jack}},\ }\href@noop {} {\bibfield  {journal} {\bibinfo  {journal} {Annual
  review of physical chemistry}\ }\textbf {\bibinfo {volume} {66}},\ \bibinfo
  {pages} {143} (\bibinfo {year} {2015})}\BibitemShut {NoStop}%
\bibitem [{\citenamefont {Whitelam}\ \emph {et~al.}(2009)\citenamefont
  {Whitelam}, \citenamefont {Feng}, \citenamefont {Hagan},\ and\ \citenamefont
  {Geissler}}]{whitelam2009role}%
  \BibitemOpen
  \bibfield  {author} {\bibinfo {author} {\bibfnamefont {S.}~\bibnamefont
  {Whitelam}}, \bibinfo {author} {\bibfnamefont {E.~H.}\ \bibnamefont {Feng}},
  \bibinfo {author} {\bibfnamefont {M.~F.}\ \bibnamefont {Hagan}}, \ and\
  \bibinfo {author} {\bibfnamefont {P.~L.}\ \bibnamefont {Geissler}},\
  }\href@noop {} {\bibfield  {journal} {\bibinfo  {journal} {Soft Matter}\
  }\textbf {\bibinfo {volume} {5}},\ \bibinfo {pages} {1251} (\bibinfo {year}
  {2009})}\BibitemShut {NoStop}%
\bibitem [{\citenamefont {Kuznets-Speck}\ and\ \citenamefont
  {Limmer}(2021)}]{kuznets2021dissipation}%
  \BibitemOpen
  \bibfield  {author} {\bibinfo {author} {\bibfnamefont {B.}~\bibnamefont
  {Kuznets-Speck}}\ and\ \bibinfo {author} {\bibfnamefont {D.~T.}\ \bibnamefont
  {Limmer}},\ }\href@noop {} {\bibfield  {journal} {\bibinfo  {journal}
  {Proceedings of the National Academy of Sciences}\ }\textbf {\bibinfo
  {volume} {118}} (\bibinfo {year} {2021})}\BibitemShut {NoStop}%
\bibitem [{\citenamefont {Das}\ \emph {et~al.}(2022)\citenamefont {Das},
  \citenamefont {Kuznets-Speck},\ and\ \citenamefont {Limmer}}]{das2022direct}%
  \BibitemOpen
  \bibfield  {author} {\bibinfo {author} {\bibfnamefont {A.}~\bibnamefont
  {Das}}, \bibinfo {author} {\bibfnamefont {B.}~\bibnamefont {Kuznets-Speck}},
  \ and\ \bibinfo {author} {\bibfnamefont {D.~T.}\ \bibnamefont {Limmer}},\
  }\href@noop {} {\bibfield  {journal} {\bibinfo  {journal} {Physical Review
  Letters}\ }\textbf {\bibinfo {volume} {128}},\ \bibinfo {pages} {028005}
  (\bibinfo {year} {2022})}\BibitemShut {NoStop}%
\bibitem [{\citenamefont {Whitelam}(2018)}]{whitelam2018strong}%
  \BibitemOpen
  \bibfield  {author} {\bibinfo {author} {\bibfnamefont {S.}~\bibnamefont
  {Whitelam}},\ }\href@noop {} {\bibfield  {journal} {\bibinfo  {journal} {The
  Journal of chemical physics}\ }\textbf {\bibinfo {volume} {149}},\ \bibinfo
  {pages} {104902} (\bibinfo {year} {2018})}\BibitemShut {NoStop}%
\bibitem [{\citenamefont {Nguyen}\ \emph {et~al.}(2021)\citenamefont {Nguyen},
  \citenamefont {Qiu},\ and\ \citenamefont
  {Vaikuntanathan}}]{nguyen2021organization}%
  \BibitemOpen
  \bibfield  {author} {\bibinfo {author} {\bibfnamefont {M.}~\bibnamefont
  {Nguyen}}, \bibinfo {author} {\bibfnamefont {Y.}~\bibnamefont {Qiu}}, \ and\
  \bibinfo {author} {\bibfnamefont {S.}~\bibnamefont {Vaikuntanathan}},\
  }\href@noop {} {\bibfield  {journal} {\bibinfo  {journal} {Annual Review of
  Condensed Matter Physics}\ }\textbf {\bibinfo {volume} {12}} (\bibinfo {year}
  {2021})}\BibitemShut {NoStop}%
\bibitem [{\citenamefont {Charbonneau}\ and\ \citenamefont
  {Reichman}(2007)}]{charbonneau2007phase}%
  \BibitemOpen
  \bibfield  {author} {\bibinfo {author} {\bibfnamefont {P.}~\bibnamefont
  {Charbonneau}}\ and\ \bibinfo {author} {\bibfnamefont {D.}~\bibnamefont
  {Reichman}},\ }\href@noop {} {\bibfield  {journal} {\bibinfo  {journal}
  {Physical Review E}\ }\textbf {\bibinfo {volume} {75}},\ \bibinfo {pages}
  {050401} (\bibinfo {year} {2007})}\BibitemShut {NoStop}%
\bibitem [{\citenamefont {Chremos}\ \emph {et~al.}(2010)\citenamefont
  {Chremos}, \citenamefont {Margaritis},\ and\ \citenamefont
  {Panagiotopoulos}}]{chremos2010ultra}%
  \BibitemOpen
  \bibfield  {author} {\bibinfo {author} {\bibfnamefont {A.}~\bibnamefont
  {Chremos}}, \bibinfo {author} {\bibfnamefont {K.}~\bibnamefont {Margaritis}},
  \ and\ \bibinfo {author} {\bibfnamefont {A.~Z.}\ \bibnamefont
  {Panagiotopoulos}},\ }\href@noop {} {\bibfield  {journal} {\bibinfo
  {journal} {Soft Matter}\ }\textbf {\bibinfo {volume} {6}},\ \bibinfo {pages}
  {3588} (\bibinfo {year} {2010})}\BibitemShut {NoStop}%
\bibitem [{\citenamefont {Stenhammar}\ \emph {et~al.}(2013)\citenamefont
  {Stenhammar}, \citenamefont {Tiribocchi}, \citenamefont {Allen},
  \citenamefont {Marenduzzo},\ and\ \citenamefont
  {Cates}}]{stenhammar2013continuum}%
  \BibitemOpen
  \bibfield  {author} {\bibinfo {author} {\bibfnamefont {J.}~\bibnamefont
  {Stenhammar}}, \bibinfo {author} {\bibfnamefont {A.}~\bibnamefont
  {Tiribocchi}}, \bibinfo {author} {\bibfnamefont {R.~J.}\ \bibnamefont
  {Allen}}, \bibinfo {author} {\bibfnamefont {D.}~\bibnamefont {Marenduzzo}}, \
  and\ \bibinfo {author} {\bibfnamefont {M.~E.}\ \bibnamefont {Cates}},\
  }\href@noop {} {\bibfield  {journal} {\bibinfo  {journal} {Physical review
  letters}\ }\textbf {\bibinfo {volume} {111}},\ \bibinfo {pages} {145702}
  (\bibinfo {year} {2013})}\BibitemShut {NoStop}%
\bibitem [{\citenamefont {Wang}\ \emph {et~al.}(2009)\citenamefont {Wang},
  \citenamefont {Xu},\ and\ \citenamefont {Zhang}}]{wang2009tuning}%
  \BibitemOpen
  \bibfield  {author} {\bibinfo {author} {\bibfnamefont {Y.}~\bibnamefont
  {Wang}}, \bibinfo {author} {\bibfnamefont {H.}~\bibnamefont {Xu}}, \ and\
  \bibinfo {author} {\bibfnamefont {X.}~\bibnamefont {Zhang}},\ }\href@noop {}
  {\bibfield  {journal} {\bibinfo  {journal} {Advanced Materials}\ }\textbf
  {\bibinfo {volume} {21}},\ \bibinfo {pages} {2849} (\bibinfo {year}
  {2009})}\BibitemShut {NoStop}%
\bibitem [{\citenamefont {Atkin-Smith}\ and\ \citenamefont
  {Poon}(2017)}]{atkin2017disassembly}%
  \BibitemOpen
  \bibfield  {author} {\bibinfo {author} {\bibfnamefont {G.~K.}\ \bibnamefont
  {Atkin-Smith}}\ and\ \bibinfo {author} {\bibfnamefont {I.~K.}\ \bibnamefont
  {Poon}},\ }\href@noop {} {\bibfield  {journal} {\bibinfo  {journal} {Trends
  in cell biology}\ }\textbf {\bibinfo {volume} {27}},\ \bibinfo {pages} {151}
  (\bibinfo {year} {2017})}\BibitemShut {NoStop}%
\bibitem [{\citenamefont {Grzybowski}\ \emph {et~al.}(2017)\citenamefont
  {Grzybowski}, \citenamefont {Fitzner}, \citenamefont {Paczesny},\ and\
  \citenamefont {Granick}}]{grzybowski2017dynamic}%
  \BibitemOpen
  \bibfield  {author} {\bibinfo {author} {\bibfnamefont {B.~A.}\ \bibnamefont
  {Grzybowski}}, \bibinfo {author} {\bibfnamefont {K.}~\bibnamefont {Fitzner}},
  \bibinfo {author} {\bibfnamefont {J.}~\bibnamefont {Paczesny}}, \ and\
  \bibinfo {author} {\bibfnamefont {S.}~\bibnamefont {Granick}},\ }\href@noop
  {} {\bibfield  {journal} {\bibinfo  {journal} {Chemical Society Reviews}\
  }\textbf {\bibinfo {volume} {46}},\ \bibinfo {pages} {5647} (\bibinfo {year}
  {2017})}\BibitemShut {NoStop}%
\bibitem [{\citenamefont {Murugan}\ \emph {et~al.}(2012)\citenamefont
  {Murugan}, \citenamefont {Huse},\ and\ \citenamefont
  {Leibler}}]{murugan2012speed}%
  \BibitemOpen
  \bibfield  {author} {\bibinfo {author} {\bibfnamefont {A.}~\bibnamefont
  {Murugan}}, \bibinfo {author} {\bibfnamefont {D.~A.}\ \bibnamefont {Huse}}, \
  and\ \bibinfo {author} {\bibfnamefont {S.}~\bibnamefont {Leibler}},\
  }\href@noop {} {\bibfield  {journal} {\bibinfo  {journal} {Proceedings of the
  National Academy of Sciences}\ }\textbf {\bibinfo {volume} {109}},\ \bibinfo
  {pages} {12034} (\bibinfo {year} {2012})}\BibitemShut {NoStop}%
\bibitem [{\citenamefont {Knoch}\ and\ \citenamefont
  {Speck}(2017)}]{knoch2017nonequilibrium}%
  \BibitemOpen
  \bibfield  {author} {\bibinfo {author} {\bibfnamefont {F.}~\bibnamefont
  {Knoch}}\ and\ \bibinfo {author} {\bibfnamefont {T.}~\bibnamefont {Speck}},\
  }\href@noop {} {\bibfield  {journal} {\bibinfo  {journal} {Physical Review
  E}\ }\textbf {\bibinfo {volume} {95}},\ \bibinfo {pages} {012503} (\bibinfo
  {year} {2017})}\BibitemShut {NoStop}%
\bibitem [{\citenamefont {Albaugh}\ and\ \citenamefont
  {Gingrich}(2022)}]{albaugh2022simulating}%
  \BibitemOpen
  \bibfield  {author} {\bibinfo {author} {\bibfnamefont {A.}~\bibnamefont
  {Albaugh}}\ and\ \bibinfo {author} {\bibfnamefont {T.~R.}\ \bibnamefont
  {Gingrich}},\ }\href@noop {} {\bibfield  {journal} {\bibinfo  {journal}
  {Nature communications}\ }\textbf {\bibinfo {volume} {13}},\ \bibinfo {pages}
  {1} (\bibinfo {year} {2022})}\BibitemShut {NoStop}%
\bibitem [{\citenamefont {Seifert}(2012)}]{seifert2012stochastic}%
  \BibitemOpen
  \bibfield  {author} {\bibinfo {author} {\bibfnamefont {U.}~\bibnamefont
  {Seifert}},\ }\href@noop {} {\bibfield  {journal} {\bibinfo  {journal} {Rep.
  Prog. Phys.}\ }\textbf {\bibinfo {volume} {75}},\ \bibinfo {pages} {126001}
  (\bibinfo {year} {2012})}\BibitemShut {NoStop}%
\bibitem [{\citenamefont {Das}\ and\ \citenamefont
  {Limmer}(2021)}]{das2021variational}%
  \BibitemOpen
  \bibfield  {author} {\bibinfo {author} {\bibfnamefont {A.}~\bibnamefont
  {Das}}\ and\ \bibinfo {author} {\bibfnamefont {D.~T.}\ \bibnamefont
  {Limmer}},\ }\href@noop {} {\bibfield  {journal} {\bibinfo  {journal} {The
  Journal of chemical physics}\ }\textbf {\bibinfo {volume} {154}},\ \bibinfo
  {pages} {014107} (\bibinfo {year} {2021})}\BibitemShut {NoStop}%
\bibitem [{\citenamefont {Das}\ and\ \citenamefont
  {Limmer}(2019)}]{das2019variational}%
  \BibitemOpen
  \bibfield  {author} {\bibinfo {author} {\bibfnamefont {A.}~\bibnamefont
  {Das}}\ and\ \bibinfo {author} {\bibfnamefont {D.~T.}\ \bibnamefont
  {Limmer}},\ }\href@noop {} {\bibfield  {journal} {\bibinfo  {journal} {The
  Journal of chemical physics}\ }\textbf {\bibinfo {volume} {151}},\ \bibinfo
  {pages} {244123} (\bibinfo {year} {2019})}\BibitemShut {NoStop}%
\bibitem [{\citenamefont {Hagan}\ and\ \citenamefont
  {Chandler}(2006)}]{hagan2006dynamic}%
  \BibitemOpen
  \bibfield  {author} {\bibinfo {author} {\bibfnamefont {M.~F.}\ \bibnamefont
  {Hagan}}\ and\ \bibinfo {author} {\bibfnamefont {D.}~\bibnamefont
  {Chandler}},\ }\href@noop {} {\bibfield  {journal} {\bibinfo  {journal}
  {Biophysical journal}\ }\textbf {\bibinfo {volume} {91}},\ \bibinfo {pages}
  {42} (\bibinfo {year} {2006})}\BibitemShut {NoStop}%
\bibitem [{\citenamefont {Newton}\ \emph {et~al.}(2015)\citenamefont {Newton},
  \citenamefont {Groenewold}, \citenamefont {Kegel},\ and\ \citenamefont
  {Bolhuis}}]{newton2015rotational}%
  \BibitemOpen
  \bibfield  {author} {\bibinfo {author} {\bibfnamefont {A.~C.}\ \bibnamefont
  {Newton}}, \bibinfo {author} {\bibfnamefont {J.}~\bibnamefont {Groenewold}},
  \bibinfo {author} {\bibfnamefont {W.~K.}\ \bibnamefont {Kegel}}, \ and\
  \bibinfo {author} {\bibfnamefont {P.~G.}\ \bibnamefont {Bolhuis}},\
  }\href@noop {} {\bibfield  {journal} {\bibinfo  {journal} {Proceedings of the
  National Academy of Sciences}\ }\textbf {\bibinfo {volume} {112}},\ \bibinfo
  {pages} {15308} (\bibinfo {year} {2015})}\BibitemShut {NoStop}%
\bibitem [{\citenamefont {Jack}\ \emph {et~al.}(2007)\citenamefont {Jack},
  \citenamefont {Hagan},\ and\ \citenamefont {Chandler}}]{jack2007fluctuation}%
  \BibitemOpen
  \bibfield  {author} {\bibinfo {author} {\bibfnamefont {R.~L.}\ \bibnamefont
  {Jack}}, \bibinfo {author} {\bibfnamefont {M.~F.}\ \bibnamefont {Hagan}}, \
  and\ \bibinfo {author} {\bibfnamefont {D.}~\bibnamefont {Chandler}},\
  }\href@noop {} {\bibfield  {journal} {\bibinfo  {journal} {Physical Review
  E}\ }\textbf {\bibinfo {volume} {76}},\ \bibinfo {pages} {021119} (\bibinfo
  {year} {2007})}\BibitemShut {NoStop}%
\bibitem [{\citenamefont {Goodrich}\ \emph {et~al.}(2021)\citenamefont
  {Goodrich}, \citenamefont {King}, \citenamefont {Schoenholz}, \citenamefont
  {Cubuk},\ and\ \citenamefont {Brenner}}]{goodrich2020self}%
  \BibitemOpen
  \bibfield  {author} {\bibinfo {author} {\bibfnamefont {C.~P.}\ \bibnamefont
  {Goodrich}}, \bibinfo {author} {\bibfnamefont {E.~M.}\ \bibnamefont {King}},
  \bibinfo {author} {\bibfnamefont {S.~S.}\ \bibnamefont {Schoenholz}},
  \bibinfo {author} {\bibfnamefont {E.~D.}\ \bibnamefont {Cubuk}}, \ and\
  \bibinfo {author} {\bibfnamefont {M.~P.}\ \bibnamefont {Brenner}},\
  }\href@noop {} {\bibfield  {journal} {\bibinfo  {journal} {Proceedings of the
  National Academy of Sciences}\ }\textbf {\bibinfo {volume} {118}} (\bibinfo
  {year} {2021})}\BibitemShut {NoStop}%
\bibitem [{\citenamefont {Miskin}\ \emph {et~al.}(2016)\citenamefont {Miskin},
  \citenamefont {Khaira}, \citenamefont {de~Pablo},\ and\ \citenamefont
  {Jaeger}}]{miskin2016turning}%
  \BibitemOpen
  \bibfield  {author} {\bibinfo {author} {\bibfnamefont {M.~Z.}\ \bibnamefont
  {Miskin}}, \bibinfo {author} {\bibfnamefont {G.}~\bibnamefont {Khaira}},
  \bibinfo {author} {\bibfnamefont {J.~J.}\ \bibnamefont {de~Pablo}}, \ and\
  \bibinfo {author} {\bibfnamefont {H.~M.}\ \bibnamefont {Jaeger}},\
  }\href@noop {} {\bibfield  {journal} {\bibinfo  {journal} {Proc. Natl. Acad.
  Sci. USA}\ }\textbf {\bibinfo {volume} {113}},\ \bibinfo {pages} {34}
  (\bibinfo {year} {2016})}\BibitemShut {NoStop}%
\bibitem [{\citenamefont {Weeks}\ \emph {et~al.}(1971)\citenamefont {Weeks},
  \citenamefont {Chandler},\ and\ \citenamefont {Andersen}}]{weeks1971role}%
  \BibitemOpen
  \bibfield  {author} {\bibinfo {author} {\bibfnamefont {J.~D.}\ \bibnamefont
  {Weeks}}, \bibinfo {author} {\bibfnamefont {D.}~\bibnamefont {Chandler}}, \
  and\ \bibinfo {author} {\bibfnamefont {H.~C.}\ \bibnamefont {Andersen}},\
  }\href@noop {} {\bibfield  {journal} {\bibinfo  {journal} {The Journal of
  chemical physics}\ }\textbf {\bibinfo {volume} {54}},\ \bibinfo {pages}
  {5237} (\bibinfo {year} {1971})}\BibitemShut {NoStop}%
\bibitem [{\citenamefont {Rogers}\ and\ \citenamefont
  {Crocker}(2011)}]{rogers2011direct}%
  \BibitemOpen
  \bibfield  {author} {\bibinfo {author} {\bibfnamefont {W.~B.}\ \bibnamefont
  {Rogers}}\ and\ \bibinfo {author} {\bibfnamefont {J.~C.}\ \bibnamefont
  {Crocker}},\ }\href@noop {} {\bibfield  {journal} {\bibinfo  {journal}
  {Proceedings of the National Academy of Sciences}\ }\textbf {\bibinfo
  {volume} {108}},\ \bibinfo {pages} {15687} (\bibinfo {year}
  {2011})}\BibitemShut {NoStop}%
\bibitem [{\citenamefont {Manoharan}(2015)}]{manoharan2015colloidal}%
  \BibitemOpen
  \bibfield  {author} {\bibinfo {author} {\bibfnamefont {V.~N.}\ \bibnamefont
  {Manoharan}},\ }\href@noop {} {\bibfield  {journal} {\bibinfo  {journal}
  {Science}\ }\textbf {\bibinfo {volume} {349}} (\bibinfo {year}
  {2015})}\BibitemShut {NoStop}%
\bibitem [{\citenamefont {Meng}\ \emph {et~al.}(2010)\citenamefont {Meng},
  \citenamefont {Arkus}, \citenamefont {Brenner},\ and\ \citenamefont
  {Manoharan}}]{meng2010free}%
  \BibitemOpen
  \bibfield  {author} {\bibinfo {author} {\bibfnamefont {G.}~\bibnamefont
  {Meng}}, \bibinfo {author} {\bibfnamefont {N.}~\bibnamefont {Arkus}},
  \bibinfo {author} {\bibfnamefont {M.~P.}\ \bibnamefont {Brenner}}, \ and\
  \bibinfo {author} {\bibfnamefont {V.~N.}\ \bibnamefont {Manoharan}},\
  }\href@noop {} {\bibfield  {journal} {\bibinfo  {journal} {Science}\ }\textbf
  {\bibinfo {volume} {327}},\ \bibinfo {pages} {560} (\bibinfo {year}
  {2010})}\BibitemShut {NoStop}%
\bibitem [{\citenamefont {Zeravcic}\ \emph {et~al.}(2014)\citenamefont
  {Zeravcic}, \citenamefont {Manoharan},\ and\ \citenamefont
  {Brenner}}]{zeravcic2014size}%
  \BibitemOpen
  \bibfield  {author} {\bibinfo {author} {\bibfnamefont {Z.}~\bibnamefont
  {Zeravcic}}, \bibinfo {author} {\bibfnamefont {V.~N.}\ \bibnamefont
  {Manoharan}}, \ and\ \bibinfo {author} {\bibfnamefont {M.~P.}\ \bibnamefont
  {Brenner}},\ }\href@noop {} {\bibfield  {journal} {\bibinfo  {journal}
  {Proceedings of the National Academy of Sciences}\ }\textbf {\bibinfo
  {volume} {111}},\ \bibinfo {pages} {15918} (\bibinfo {year}
  {2014})}\BibitemShut {NoStop}%
\bibitem [{\citenamefont {Chandler}(1978)}]{chandler1978statistical}%
  \BibitemOpen
  \bibfield  {author} {\bibinfo {author} {\bibfnamefont {D.}~\bibnamefont
  {Chandler}},\ }\href@noop {} {\bibfield  {journal} {\bibinfo  {journal} {The
  Journal of Chemical Physics}\ }\textbf {\bibinfo {volume} {68}},\ \bibinfo
  {pages} {2959} (\bibinfo {year} {1978})}\BibitemShut {NoStop}%
\bibitem [{\citenamefont {Onsager}\ and\ \citenamefont
  {Machlup}(1953)}]{Onsager_Machlup}%
  \BibitemOpen
  \bibfield  {author} {\bibinfo {author} {\bibfnamefont {L.}~\bibnamefont
  {Onsager}}\ and\ \bibinfo {author} {\bibfnamefont {S.}~\bibnamefont
  {Machlup}},\ }\href {\doibase https://doi.org/10.1103/PhysRev.91.1505}
  {\bibfield  {journal} {\bibinfo  {journal} {Phys. Rev.}\ }\textbf {\bibinfo
  {volume} {91}},\ \bibinfo {pages} {1505} (\bibinfo {year}
  {1953})}\BibitemShut {NoStop}%
\bibitem [{\citenamefont {Warren}\ and\ \citenamefont
  {Allen}(2012)}]{warren2012malliavin}%
  \BibitemOpen
  \bibfield  {author} {\bibinfo {author} {\bibfnamefont {P.~B.}\ \bibnamefont
  {Warren}}\ and\ \bibinfo {author} {\bibfnamefont {R.~J.}\ \bibnamefont
  {Allen}},\ }\href@noop {} {\bibfield  {journal} {\bibinfo  {journal} {Phys.
  Rev. Lett.}\ }\textbf {\bibinfo {volume} {109}},\ \bibinfo {pages} {250601}
  (\bibinfo {year} {2012})}\BibitemShut {NoStop}%
\bibitem [{\citenamefont {Hormoz}\ and\ \citenamefont
  {Brenner}(2011)}]{hormoz2011design}%
  \BibitemOpen
  \bibfield  {author} {\bibinfo {author} {\bibfnamefont {S.}~\bibnamefont
  {Hormoz}}\ and\ \bibinfo {author} {\bibfnamefont {M.~P.}\ \bibnamefont
  {Brenner}},\ }\href@noop {} {\bibfield  {journal} {\bibinfo  {journal}
  {Proceedings of the National Academy of Sciences}\ }\textbf {\bibinfo
  {volume} {108}},\ \bibinfo {pages} {5193} (\bibinfo {year}
  {2011})}\BibitemShut {NoStop}%
\bibitem [{\citenamefont {Dou}\ \emph {et~al.}(2017)\citenamefont {Dou},
  \citenamefont {Wang}, \citenamefont {Jin}, \citenamefont {Yu}, \citenamefont
  {Zhou},\ and\ \citenamefont {Shui}}]{dou2017review}%
  \BibitemOpen
  \bibfield  {author} {\bibinfo {author} {\bibfnamefont {Y.}~\bibnamefont
  {Dou}}, \bibinfo {author} {\bibfnamefont {B.}~\bibnamefont {Wang}}, \bibinfo
  {author} {\bibfnamefont {M.}~\bibnamefont {Jin}}, \bibinfo {author}
  {\bibfnamefont {Y.}~\bibnamefont {Yu}}, \bibinfo {author} {\bibfnamefont
  {G.}~\bibnamefont {Zhou}}, \ and\ \bibinfo {author} {\bibfnamefont
  {L.}~\bibnamefont {Shui}},\ }\href@noop {} {\bibfield  {journal} {\bibinfo
  {journal} {Journal of Micromechanics and Microengineering}\ }\textbf
  {\bibinfo {volume} {27}},\ \bibinfo {pages} {113002} (\bibinfo {year}
  {2017})}\BibitemShut {NoStop}%
\bibitem [{\citenamefont {Richard}\ and\ \citenamefont
  {Speck}(2015)}]{richard2015role}%
  \BibitemOpen
  \bibfield  {author} {\bibinfo {author} {\bibfnamefont {D.}~\bibnamefont
  {Richard}}\ and\ \bibinfo {author} {\bibfnamefont {T.}~\bibnamefont
  {Speck}},\ }\href@noop {} {\bibfield  {journal} {\bibinfo  {journal}
  {Scientific reports}\ }\textbf {\bibinfo {volume} {5}},\ \bibinfo {pages} {1}
  (\bibinfo {year} {2015})}\BibitemShut {NoStop}%
\bibitem [{\citenamefont {Gardiner}\ \emph {et~al.}(1985)\citenamefont
  {Gardiner} \emph {et~al.}}]{gardiner1985handbook}%
  \BibitemOpen
  \bibfield  {author} {\bibinfo {author} {\bibfnamefont {C.~W.}\ \bibnamefont
  {Gardiner}} \emph {et~al.},\ }\href@noop {} {\emph {\bibinfo {title}
  {Handbook of stochastic methods}}},\ Vol.~\bibinfo {volume} {3}\ (\bibinfo
  {publisher} {springer Berlin},\ \bibinfo {year} {1985})\BibitemShut {NoStop}%
\bibitem [{\citenamefont {Schneider}\ \emph {et~al.}(2007)\citenamefont
  {Schneider}, \citenamefont {Nuschele}, \citenamefont {Wixforth},
  \citenamefont {Gorzelanny}, \citenamefont {Alexander-Katz}, \citenamefont
  {Netz},\ and\ \citenamefont {Schneider}}]{schneider2007shear}%
  \BibitemOpen
  \bibfield  {author} {\bibinfo {author} {\bibfnamefont {S.}~\bibnamefont
  {Schneider}}, \bibinfo {author} {\bibfnamefont {S.}~\bibnamefont {Nuschele}},
  \bibinfo {author} {\bibfnamefont {A.}~\bibnamefont {Wixforth}}, \bibinfo
  {author} {\bibfnamefont {C.}~\bibnamefont {Gorzelanny}}, \bibinfo {author}
  {\bibfnamefont {A.}~\bibnamefont {Alexander-Katz}}, \bibinfo {author}
  {\bibfnamefont {R.}~\bibnamefont {Netz}}, \ and\ \bibinfo {author}
  {\bibfnamefont {M.~F.}\ \bibnamefont {Schneider}},\ }\href@noop {} {\bibfield
   {journal} {\bibinfo  {journal} {Proceedings of the National Academy of
  Sciences}\ }\textbf {\bibinfo {volume} {104}},\ \bibinfo {pages} {7899}
  (\bibinfo {year} {2007})}\BibitemShut {NoStop}%
\bibitem [{\citenamefont {Vermant}\ and\ \citenamefont
  {Solomon}(2005)}]{vermant2005flow}%
  \BibitemOpen
  \bibfield  {author} {\bibinfo {author} {\bibfnamefont {J.}~\bibnamefont
  {Vermant}}\ and\ \bibinfo {author} {\bibfnamefont {M.~J.}\ \bibnamefont
  {Solomon}},\ }\href@noop {} {\bibfield  {journal} {\bibinfo  {journal}
  {Journal of Physics: Condensed Matter}\ }\textbf {\bibinfo {volume} {17}},\
  \bibinfo {pages} {R187} (\bibinfo {year} {2005})}\BibitemShut {NoStop}%
\bibitem [{\citenamefont {Toyabe}\ \emph {et~al.}(2007)\citenamefont {Toyabe},
  \citenamefont {Jiang}, \citenamefont {Nakamura}, \citenamefont {Murayama},\
  and\ \citenamefont {Sano}}]{toyabe2007experimental}%
  \BibitemOpen
  \bibfield  {author} {\bibinfo {author} {\bibfnamefont {S.}~\bibnamefont
  {Toyabe}}, \bibinfo {author} {\bibfnamefont {H.-R.}\ \bibnamefont {Jiang}},
  \bibinfo {author} {\bibfnamefont {T.}~\bibnamefont {Nakamura}}, \bibinfo
  {author} {\bibfnamefont {Y.}~\bibnamefont {Murayama}}, \ and\ \bibinfo
  {author} {\bibfnamefont {M.}~\bibnamefont {Sano}},\ }\href@noop {} {\bibfield
   {journal} {\bibinfo  {journal} {Physical Review E}\ }\textbf {\bibinfo
  {volume} {75}},\ \bibinfo {pages} {011122} (\bibinfo {year}
  {2007})}\BibitemShut {NoStop}%
\bibitem [{\citenamefont {Wu}\ \emph {et~al.}(2006)\citenamefont {Wu},
  \citenamefont {Register},\ and\ \citenamefont {Chaikin}}]{wu2006shear}%
  \BibitemOpen
  \bibfield  {author} {\bibinfo {author} {\bibfnamefont {M.~W.}\ \bibnamefont
  {Wu}}, \bibinfo {author} {\bibfnamefont {R.~A.}\ \bibnamefont {Register}}, \
  and\ \bibinfo {author} {\bibfnamefont {P.~M.}\ \bibnamefont {Chaikin}},\
  }\href@noop {} {\bibfield  {journal} {\bibinfo  {journal} {Physical Review
  E}\ }\textbf {\bibinfo {volume} {74}},\ \bibinfo {pages} {040801} (\bibinfo
  {year} {2006})}\BibitemShut {NoStop}%
\bibitem [{\citenamefont {Alexander-Katz}\ \emph {et~al.}(2006)\citenamefont
  {Alexander-Katz}, \citenamefont {Schneider}, \citenamefont {Schneider},
  \citenamefont {Wixforth},\ and\ \citenamefont {Netz}}]{alexander2006shear}%
  \BibitemOpen
  \bibfield  {author} {\bibinfo {author} {\bibfnamefont {A.}~\bibnamefont
  {Alexander-Katz}}, \bibinfo {author} {\bibfnamefont {M.}~\bibnamefont
  {Schneider}}, \bibinfo {author} {\bibfnamefont {S.}~\bibnamefont
  {Schneider}}, \bibinfo {author} {\bibfnamefont {A.}~\bibnamefont {Wixforth}},
  \ and\ \bibinfo {author} {\bibfnamefont {R.}~\bibnamefont {Netz}},\
  }\href@noop {} {\bibfield  {journal} {\bibinfo  {journal} {Physical review
  letters}\ }\textbf {\bibinfo {volume} {97}},\ \bibinfo {pages} {138101}
  (\bibinfo {year} {2006})}\BibitemShut {NoStop}%
\bibitem [{\citenamefont {Hill}\ \emph {et~al.}(2006)\citenamefont {Hill},
  \citenamefont {Krebs}, \citenamefont {Goodall}, \citenamefont {Howlett},\
  and\ \citenamefont {Dunstan}}]{hill2006shear}%
  \BibitemOpen
  \bibfield  {author} {\bibinfo {author} {\bibfnamefont {E.~K.}\ \bibnamefont
  {Hill}}, \bibinfo {author} {\bibfnamefont {B.}~\bibnamefont {Krebs}},
  \bibinfo {author} {\bibfnamefont {D.~G.}\ \bibnamefont {Goodall}}, \bibinfo
  {author} {\bibfnamefont {G.~J.}\ \bibnamefont {Howlett}}, \ and\ \bibinfo
  {author} {\bibfnamefont {D.~E.}\ \bibnamefont {Dunstan}},\ }\href@noop {}
  {\bibfield  {journal} {\bibinfo  {journal} {Biomacromolecules}\ }\textbf
  {\bibinfo {volume} {7}},\ \bibinfo {pages} {10} (\bibinfo {year}
  {2006})}\BibitemShut {NoStop}%
\bibitem [{\citenamefont {Liphardt}\ \emph {et~al.}(2002)\citenamefont
  {Liphardt}, \citenamefont {Dumont}, \citenamefont {Smith}, \citenamefont
  {Tinoco~Jr},\ and\ \citenamefont {Bustamante}}]{liphardt2002equilibrium}%
  \BibitemOpen
  \bibfield  {author} {\bibinfo {author} {\bibfnamefont {J.}~\bibnamefont
  {Liphardt}}, \bibinfo {author} {\bibfnamefont {S.}~\bibnamefont {Dumont}},
  \bibinfo {author} {\bibfnamefont {S.~B.}\ \bibnamefont {Smith}}, \bibinfo
  {author} {\bibfnamefont {I.}~\bibnamefont {Tinoco~Jr}}, \ and\ \bibinfo
  {author} {\bibfnamefont {C.}~\bibnamefont {Bustamante}},\ }\href@noop {}
  {\bibfield  {journal} {\bibinfo  {journal} {Science}\ }\textbf {\bibinfo
  {volume} {296}},\ \bibinfo {pages} {1832} (\bibinfo {year}
  {2002})}\BibitemShut {NoStop}%
\bibitem [{\citenamefont {Zhuang}\ and\ \citenamefont
  {Charbonneau}(2016)}]{zhuang2016recent}%
  \BibitemOpen
  \bibfield  {author} {\bibinfo {author} {\bibfnamefont {Y.}~\bibnamefont
  {Zhuang}}\ and\ \bibinfo {author} {\bibfnamefont {P.}~\bibnamefont
  {Charbonneau}},\ }\href@noop {} {\bibfield  {journal} {\bibinfo  {journal}
  {The Journal of Physical Chemistry B}\ }\textbf {\bibinfo {volume} {120}},\
  \bibinfo {pages} {7775} (\bibinfo {year} {2016})}\BibitemShut {NoStop}%
\bibitem [{\citenamefont {Hagan}\ and\ \citenamefont
  {Grason}(2021)}]{hagan2021equilibrium}%
  \BibitemOpen
  \bibfield  {author} {\bibinfo {author} {\bibfnamefont {M.~F.}\ \bibnamefont
  {Hagan}}\ and\ \bibinfo {author} {\bibfnamefont {G.~M.}\ \bibnamefont
  {Grason}},\ }\href@noop {} {\bibfield  {journal} {\bibinfo  {journal}
  {Reviews of Modern Physics}\ }\textbf {\bibinfo {volume} {93}},\ \bibinfo
  {pages} {025008} (\bibinfo {year} {2021})}\BibitemShut {NoStop}%
\bibitem [{\citenamefont {Caspar}\ and\ \citenamefont
  {Klug}(1962)}]{caspar1962physical}%
  \BibitemOpen
  \bibfield  {author} {\bibinfo {author} {\bibfnamefont {D.~L.}\ \bibnamefont
  {Caspar}}\ and\ \bibinfo {author} {\bibfnamefont {A.}~\bibnamefont {Klug}},\
  }in\ \href@noop {} {\emph {\bibinfo {booktitle} {Cold Spring Harbor symposia
  on quantitative biology}}},\ Vol.~\bibinfo {volume} {27}\ (\bibinfo
  {organization} {Cold Spring Harbor Laboratory Press},\ \bibinfo {year}
  {1962})\ pp.\ \bibinfo {pages} {1--24}\BibitemShut {NoStop}%
\bibitem [{\citenamefont {Tanaka}\ \emph {et~al.}(2008)\citenamefont {Tanaka},
  \citenamefont {Kerfeld}, \citenamefont {Sawaya}, \citenamefont {Cai},
  \citenamefont {Heinhorst}, \citenamefont {Cannon},\ and\ \citenamefont
  {Yeates}}]{tanaka2008atomic}%
  \BibitemOpen
  \bibfield  {author} {\bibinfo {author} {\bibfnamefont {S.}~\bibnamefont
  {Tanaka}}, \bibinfo {author} {\bibfnamefont {C.~A.}\ \bibnamefont {Kerfeld}},
  \bibinfo {author} {\bibfnamefont {M.~R.}\ \bibnamefont {Sawaya}}, \bibinfo
  {author} {\bibfnamefont {F.}~\bibnamefont {Cai}}, \bibinfo {author}
  {\bibfnamefont {S.}~\bibnamefont {Heinhorst}}, \bibinfo {author}
  {\bibfnamefont {G.~C.}\ \bibnamefont {Cannon}}, \ and\ \bibinfo {author}
  {\bibfnamefont {T.~O.}\ \bibnamefont {Yeates}},\ }\href@noop {} {\bibfield
  {journal} {\bibinfo  {journal} {science}\ }\textbf {\bibinfo {volume}
  {319}},\ \bibinfo {pages} {1083} (\bibinfo {year} {2008})}\BibitemShut
  {NoStop}%
\bibitem [{\citenamefont {Courchaine}\ \emph {et~al.}(2016)\citenamefont
  {Courchaine}, \citenamefont {Lu},\ and\ \citenamefont
  {Neugebauer}}]{courchaine2016droplet}%
  \BibitemOpen
  \bibfield  {author} {\bibinfo {author} {\bibfnamefont {E.~M.}\ \bibnamefont
  {Courchaine}}, \bibinfo {author} {\bibfnamefont {A.}~\bibnamefont {Lu}}, \
  and\ \bibinfo {author} {\bibfnamefont {K.~M.}\ \bibnamefont {Neugebauer}},\
  }\href@noop {} {\bibfield  {journal} {\bibinfo  {journal} {The EMBO journal}\
  }\textbf {\bibinfo {volume} {35}},\ \bibinfo {pages} {1603} (\bibinfo {year}
  {2016})}\BibitemShut {NoStop}%
\bibitem [{\citenamefont {Zwicker}\ \emph {et~al.}(2015)\citenamefont
  {Zwicker}, \citenamefont {Hyman},\ and\ \citenamefont
  {J{\"u}licher}}]{zwicker2015suppression}%
  \BibitemOpen
  \bibfield  {author} {\bibinfo {author} {\bibfnamefont {D.}~\bibnamefont
  {Zwicker}}, \bibinfo {author} {\bibfnamefont {A.~A.}\ \bibnamefont {Hyman}},
  \ and\ \bibinfo {author} {\bibfnamefont {F.}~\bibnamefont {J{\"u}licher}},\
  }\href@noop {} {\bibfield  {journal} {\bibinfo  {journal} {Physical Review
  E}\ }\textbf {\bibinfo {volume} {92}},\ \bibinfo {pages} {012317} (\bibinfo
  {year} {2015})}\BibitemShut {NoStop}%
\bibitem [{\citenamefont {S{\"o}ding}\ \emph {et~al.}(2020)\citenamefont
  {S{\"o}ding}, \citenamefont {Zwicker}, \citenamefont {Sohrabi-Jahromi},
  \citenamefont {Boehning},\ and\ \citenamefont
  {Kirschbaum}}]{soding2020mechanisms}%
  \BibitemOpen
  \bibfield  {author} {\bibinfo {author} {\bibfnamefont {J.}~\bibnamefont
  {S{\"o}ding}}, \bibinfo {author} {\bibfnamefont {D.}~\bibnamefont {Zwicker}},
  \bibinfo {author} {\bibfnamefont {S.}~\bibnamefont {Sohrabi-Jahromi}},
  \bibinfo {author} {\bibfnamefont {M.}~\bibnamefont {Boehning}}, \ and\
  \bibinfo {author} {\bibfnamefont {J.}~\bibnamefont {Kirschbaum}},\
  }\href@noop {} {\bibfield  {journal} {\bibinfo  {journal} {Trends in Cell
  Biology}\ }\textbf {\bibinfo {volume} {30}},\ \bibinfo {pages} {4} (\bibinfo
  {year} {2020})}\BibitemShut {NoStop}%
\bibitem [{\citenamefont {Kirschbaum}\ and\ \citenamefont
  {Zwicker}(2021)}]{kirschbaum2021controlling}%
  \BibitemOpen
  \bibfield  {author} {\bibinfo {author} {\bibfnamefont {J.}~\bibnamefont
  {Kirschbaum}}\ and\ \bibinfo {author} {\bibfnamefont {D.}~\bibnamefont
  {Zwicker}},\ }\href@noop {} {\bibfield  {journal} {\bibinfo  {journal}
  {Journal of the Royal Society Interface}\ }\textbf {\bibinfo {volume} {18}},\
  \bibinfo {pages} {20210255} (\bibinfo {year} {2021})}\BibitemShut {NoStop}%
\bibitem [{\citenamefont {Brackley}\ \emph {et~al.}(2017)\citenamefont
  {Brackley}, \citenamefont {Liebchen}, \citenamefont {Michieletto},
  \citenamefont {Mouvet}, \citenamefont {Cook},\ and\ \citenamefont
  {Marenduzzo}}]{brackley2017ephemeral}%
  \BibitemOpen
  \bibfield  {author} {\bibinfo {author} {\bibfnamefont {C.~A.}\ \bibnamefont
  {Brackley}}, \bibinfo {author} {\bibfnamefont {B.}~\bibnamefont {Liebchen}},
  \bibinfo {author} {\bibfnamefont {D.}~\bibnamefont {Michieletto}}, \bibinfo
  {author} {\bibfnamefont {F.}~\bibnamefont {Mouvet}}, \bibinfo {author}
  {\bibfnamefont {P.~R.}\ \bibnamefont {Cook}}, \ and\ \bibinfo {author}
  {\bibfnamefont {D.}~\bibnamefont {Marenduzzo}},\ }\href@noop {} {\bibfield
  {journal} {\bibinfo  {journal} {Biophysical journal}\ }\textbf {\bibinfo
  {volume} {112}},\ \bibinfo {pages} {1085} (\bibinfo {year}
  {2017})}\BibitemShut {NoStop}%
\bibitem [{\citenamefont {Toxvaerd}\ and\ \citenamefont
  {Dyre}(2011)}]{toxvaerd2011communication}%
  \BibitemOpen
  \bibfield  {author} {\bibinfo {author} {\bibfnamefont {S.}~\bibnamefont
  {Toxvaerd}}\ and\ \bibinfo {author} {\bibfnamefont {J.~C.}\ \bibnamefont
  {Dyre}},\ }\href@noop {} {\enquote {\bibinfo {title} {Communication: Shifted
  forces in molecular dynamics},}\ } (\bibinfo {year} {2011})\BibitemShut
  {NoStop}%
\bibitem [{\citenamefont {Michaels}\ \emph {et~al.}(2020)\citenamefont
  {Michaels}, \citenamefont {{\v{S}}ari{\'c}}, \citenamefont {Curk},
  \citenamefont {Bernfur}, \citenamefont {Arosio}, \citenamefont {Meisl},
  \citenamefont {Dear}, \citenamefont {Cohen}, \citenamefont {Dobson},
  \citenamefont {Vendruscolo} \emph {et~al.}}]{michaels2020dynamics}%
  \BibitemOpen
  \bibfield  {author} {\bibinfo {author} {\bibfnamefont {T.~C.}\ \bibnamefont
  {Michaels}}, \bibinfo {author} {\bibfnamefont {A.}~\bibnamefont
  {{\v{S}}ari{\'c}}}, \bibinfo {author} {\bibfnamefont {S.}~\bibnamefont
  {Curk}}, \bibinfo {author} {\bibfnamefont {K.}~\bibnamefont {Bernfur}},
  \bibinfo {author} {\bibfnamefont {P.}~\bibnamefont {Arosio}}, \bibinfo
  {author} {\bibfnamefont {G.}~\bibnamefont {Meisl}}, \bibinfo {author}
  {\bibfnamefont {A.~J.}\ \bibnamefont {Dear}}, \bibinfo {author}
  {\bibfnamefont {S.~I.}\ \bibnamefont {Cohen}}, \bibinfo {author}
  {\bibfnamefont {C.~M.}\ \bibnamefont {Dobson}}, \bibinfo {author}
  {\bibfnamefont {M.}~\bibnamefont {Vendruscolo}},  \emph {et~al.},\
  }\href@noop {} {\bibfield  {journal} {\bibinfo  {journal} {Nature chemistry}\
  }\textbf {\bibinfo {volume} {12}},\ \bibinfo {pages} {445} (\bibinfo {year}
  {2020})}\BibitemShut {NoStop}%
\bibitem [{\citenamefont {Sourjik}\ and\ \citenamefont
  {Berg}(2004)}]{sourjik2004functional}%
  \BibitemOpen
  \bibfield  {author} {\bibinfo {author} {\bibfnamefont {V.}~\bibnamefont
  {Sourjik}}\ and\ \bibinfo {author} {\bibfnamefont {H.~C.}\ \bibnamefont
  {Berg}},\ }\href@noop {} {\bibfield  {journal} {\bibinfo  {journal} {Nature}\
  }\textbf {\bibinfo {volume} {428}},\ \bibinfo {pages} {437} (\bibinfo {year}
  {2004})}\BibitemShut {NoStop}%
\bibitem [{\citenamefont {Whitelam}\ \emph {et~al.}(2012)\citenamefont
  {Whitelam}, \citenamefont {Schulman},\ and\ \citenamefont
  {Hedges}}]{whitelam2012self}%
  \BibitemOpen
  \bibfield  {author} {\bibinfo {author} {\bibfnamefont {S.}~\bibnamefont
  {Whitelam}}, \bibinfo {author} {\bibfnamefont {R.}~\bibnamefont {Schulman}},
  \ and\ \bibinfo {author} {\bibfnamefont {L.}~\bibnamefont {Hedges}},\
  }\href@noop {} {\bibfield  {journal} {\bibinfo  {journal} {Physical review
  letters}\ }\textbf {\bibinfo {volume} {109}},\ \bibinfo {pages} {265506}
  (\bibinfo {year} {2012})}\BibitemShut {NoStop}%
\bibitem [{\citenamefont {Das}\ and\ \citenamefont
  {Limmer}(2023)}]{das_noneq_functional_design_zenodo}%
  \BibitemOpen
  \bibfield  {author} {\bibinfo {author} {\bibfnamefont {A.}~\bibnamefont
  {Das}}\ and\ \bibinfo {author} {\bibfnamefont {D.~T.}\ \bibnamefont
  {Limmer}},\ }\href {\doibase 10.5281/zenodo.8031998} {\enquote {\bibinfo
  {title} {{Nonequilibrium design strategies for functional colloidal
  assemblies}},}\ } (\bibinfo {year} {2023})\BibitemShut {NoStop}%
\end{thebibliography}
\end{document}


\beginsupplement
\title{Supporting Information: \\Nonequilibrium design strategies for functional colloidal assemblies}

\preprint{APS/123-QED}
\author{Avishek Das}
\affiliation{Department of Chemistry, University of California, Berkeley, CA, 94720, USA \looseness=-1}
\author{David T. Limmer}
\email{dlimmer@berkeley.edu}
\affiliation{Department of Chemistry, University of California, Berkeley, CA, 94720, USA \looseness=-1}
\affiliation{Chemical Sciences Division, LBNL, Berkeley, CA, 94720, USA \looseness=-1}
\affiliation{Material Sciences Division, LBNL, Berkeley, CA, 94720, USA \looseness=-1}
\affiliation{Kavli Energy NanoSciences Institute, University of California, Berkeley, CA, 94720, USA \looseness=-1}

\date{\today}

\maketitle

\thispagestyle{empty}
\onecolumngrid



\setcounter{equation}{0}
\setcounter{figure}{0}
\setcounter{table}{0}
\setcounter{page}{1}
\makeatletter
\renewcommand{\theequation}{S\arabic{equation}}
\renewcommand{\thefigure}{S\arabic{figure}}

\section*{Simulation details}
We simulate the self-assembly of colloids in a cubic box of length $L$ using an overdamped Langevin equation in a thermal bath, as
\begin{equation}\label{eq:eom}
    \gamma\dot{\mathbf{r}}_{i}=\mathbf{u}_{i}+\boldsymbol{\eta}_{i}
\end{equation}
where $\dot{\mathbf{r}}_{i}$ refers to the time-derivative of the position of the $i$-th particle, $\mathbf{u}_{i}$ is the total force and $\boldsymbol{\eta}_{i}$ are Gaussian white noise with
\begin{equation}
\langle\pmb{\eta}_{i}(t)\rangle=0 \,, \quad \langle\pmb{\eta}_{i}(t)\pmb{\eta}_{j}(t')\rangle=2\gamma k_{\mathrm{B}}T \mathbf{I}_{3}\delta_{ij} \delta(t-t')
\end{equation}
where $\mathbf{I}_{3}$ is the $3\times3$ identity matrix and the averaging operation is performed over thermal noise. In the presence of shear flow, we use Lees-Edwards periodic boundary conditions \cite{lees1972computer}. For the first part of the paper, any given nanocluster has been effectively isolated by simulating the colloids at a packing fraction of $\phi=0.01$.
The total force $\mathbf{u}_{i}$ is composed of shear flow and conservative forces,
\begin{equation}\label{eq:force}
\mathbf{u}_{i}=\mathbf{f}^{\mathrm{S}}_{i}(\mathbf{r}_{i})-\pmb{\nabla}_{i}\sum_{j\neq i} V(r_{ij})
\end{equation}
where $r_{ij}=|\mathbf{r}_{i}-\mathbf{r}_{j}|$ is the pairwise center-to-center distance.
The shear flow has the form
\begin{equation}\label{eq:shearflow}
\mathbf{f}^{\mathrm{S}}_{i}(\mathbf{r}_{i})=fz_{i}\mathbf{\hat{x}}
\end{equation}
with the shear flow rate $f$. The potential energy is composed of Weeks-Chandler-Andersen (WCA), Morse and Yukawa potentials that represent the volume-exclusion repulsion, DNA-mediated attraction and screened Coulomb repulsion respectively. The WCA pair potential has the form 
\begin{align}\label{eq:wcapair}
V_\mathrm{WCA}(r_{ij})&= 4\epsilon_{\mathrm{WCA}} \left[\left( \frac{\sigma}{r_{ij}}\right )^{12}-\left (\frac{\sigma}{r_{ij}} \right)^{6} \right] +\epsilon_{\mathrm{WCA}} \, , \quad r_{ij}<2^{1/6}\sigma \nonumber\\
&=0 \, , \quad  r_{ij}\geq2^{1/6}\sigma
\end{align}
with particle diameter $\sigma$ and energy scale $\epsilon_{\mathrm{WCA}}=10k_{\mathrm{B}}T$. The Morse potential has the form
\begin{equation}\label{eq:morse}
V_{\mathrm{Morse}}(r_{ij})=D_{ij}\left( e^{-2\alpha(r_{ij}-2^{1/6}\sigma)}-2e^{-\alpha(r_{ij}-2^{1/6}\sigma)} \right)
\end{equation}
with $D_{ij}$ being the bond energy and $\alpha=10\sigma^{-1}$ determining its width. The Morse potential has been truncated with a shifted forces approximation \cite{toxvaerd2011communication} such that both the potential and the force decay smoothly to zero at $r=2.12\sigma$. We integrate Equation \ref{eq:eom} in an Ito fashion with an Euler algorithm \cite{frenkel2019understanding} with the time step $\delta t=5\times10^{-5}t^{*}$. All stochastic integrals throughout the paper have also been performed in the Ito sense \cite{rogers2000diffusions}. For the variational optimization iterations and for sweeping over the parameter space, trajectories have been propagated upto $t=6\times10^{5}t^{*}$ for isolated nanoclusters and upto $t=5\times 10^{4}t^{*}$ for the microphase separated state. We constrain the optimization parameters to stay within pre-specified upper and lower bounds to avoid ill-behaved potential energy functions, and prohibitively slow relaxation in the steady-state simulations. These constraints are $0\leq D_{ij}/k_{\mathrm{B}}T\leq 10$, $0\leq f/f^{*}\leq50$, $0\leq \epsilon/k_{\mathrm{B}}T\leq10$ and $0.1\leq \kappa\sigma\leq10$.

\section*{Optimization algorithm}

\begin{figure}
\begin{algorithm}[H]
	\caption{Gradient Descent with Malliavin weights}\label{algo}
	\begin{algorithmic}[1]
		\State \textbf{inputs} Variational parameters $c$ for force  $\mathbf{u}^{(c)}(\mathbf{r}^{N})$
		\State \textbf{parameters} Magnitude of first optimization step $\Delta c_{0}$; lower and upper bounds on allowed values of $c$ are $c_{l}$ and $c_{m}$; total optimization steps $I$; trajectory length $t_{f}$ consisting of $J$ time steps of duration $\delta t$ each; number of trajectories $M$
		\State \textbf{initialize} Choose initial weights $c$. Force gradient learning rate $\alpha^{c}$ will be chosen during the first iteration. Define iteration variables $i$ and $j$. Define force gradients $\delta_{\mathbf{u}}^{(c)}$. Define functional form for stepwise differential increments (rewards) $\xi$ to the temporally integrated cost-function $\lambda t_{f}O-\int_{0}^{t_{f}}dt\sum_{i=1}^{N}(\mathbf{u}_{i}-\mathbf{F}_{i})^{2}/4\gamma k_{\mathrm{B}}T $.
		\State $i\gets0$
		\Repeat
		\State Generate trajectory $X$ with first-order Euler propagation in presence of the force $\mathbf{u}^{(c)}$ and wait till it relaxes into a nonequilibrium steady-state. Configurations, times, noises (with variance $2\boldsymbol{\gamma}k_{\mathrm{B}} T\delta t$), Malliavin weights, and rewards are denoted by $\mathbf{r}^{N}_{j},t_{j},\boldsymbol{\Lambda}_{j},\zeta^{(c)}(t_{j})$ and $\xi(t_{j})=\xi_{j}$ respectively. Reward contributions from stepwise increments to $O$ are derived from products of indicators $\mathbf{1}(t_{j})$ with stored information from $\tau$ times past, $\mathbf{1}(t_{j}-\tau)$. Change in Malliavin weights over the past $\Delta t$ time is denoted as $\Delta\zeta^{(c)}(t_{j})$. Steady-state averages of rewards $\xi$ and of Malliavin weight changes $\Delta \zeta^{(c)}$ are denoted with $\overline{\xi}$ and $\overline{\Delta\zeta^{(c)}}$ respectively. Steady-state average of the first term in Eq. 3 of main text is denoted as $g^{(c)}$.
		\State $j\gets0$
		\State $\{\overline{\xi},g^{(c)},\zeta^{(c)}(t_{0}),\overline{\Delta\zeta^{(c)}},\delta_{\mathbf{u}}^{(c)}\}\gets0$
		\Repeat
		\State $\xi_{j}\gets\lambda\dot{O}(t_{j},t_{j}-\tau)-(\mathbf{u}^{(c)}-\mathbf{F})^{2}/4\gamma k_{\mathrm{B}}T$
		\State $\overline{\xi}\gets \overline{\xi}+\xi_{j}$
		\State $g^{(c)}\gets g^{(c)}-[\mathbf{u}^{(c)}(t_{j})-\mathbf{F}(t_{j})]\cdot\nabla_{c}\mathbf{u}^{(c)}(t_{j})/2\gamma k_{\mathrm{B}} T$
		\State $\dot{\zeta}^{(c)}(t_{j})\gets\boldsymbol{\Lambda}_{j}\cdot\nabla_{c}\mathbf{u}^{(c)}(t_{j})/2\gamma k_{\mathrm{B}} T\delta t$
		\State $\zeta^{(c)}(t_{j+1})\gets \zeta^{(c)}(t_{j})+\delta t\dot{\zeta}^{(c)}(t_{j})$
		\State $\Delta\zeta^{(c)}(t_{j+1})\gets\zeta^{(c)}(t_{j+1})-\zeta^{(c)}(t_{j+1}-\Delta t)$
		\State $\delta_{\mathbf{u}}^{(c)}\gets\delta_{\mathbf{u}}^{(c)}+\xi_{j}\Delta\zeta^{(c)}(t_{j+1})$
		\State $\overline{\Delta\zeta^{(c)}}\gets\overline{\Delta\zeta^{(c)}}+\Delta\zeta^{(c)}(t_{j+1})$
		\State $j\gets j+1$
		\Until{$j=J$}
		\State $\overline{\xi}\gets\overline{\xi}/J$
		\State $g^{(c)}\gets g^{(c)}/J$
		\State $\overline{\Delta\zeta^{(c)}}\gets\overline{\Delta\zeta^{(c)}}/J$
		\State $\delta_{\mathbf{u}}^{(c)}\gets\delta_{\mathbf{u}}^{(c)}/J$
		\State $\delta_{\mathbf{u}}^{(c)}\gets \delta_{\mathbf{u}}^{(c)}-\overline{\xi}\cdot\overline{\Delta\zeta^{(c)}}$
		\State $\delta_{\mathbf{u}}^{(c)}\gets \delta_{\mathbf{u}}^{(c)}+g^{(c)}$
		\State average $\delta_{\mathbf{u}}^{(c)}$ over $N$ trajectories to get $\overline{\delta}_{\mathbf{u}}^{(c)}$
		\If {$i==0$}
		    \State $\alpha^{c}=\Delta c_{0}/\overline{\delta}_{\mathbf{u}}^{(c)}$
		\EndIf
		\If {$c_{l}\leq c+\alpha^{c}\overline{\delta}_{\mathbf{u}}^{(c)}\leq c_{m}$}
		    \State $c\gets c+\alpha^{c}\overline{\delta}_{\mathbf{u}}^{(c)}$
		\ElsIf {$c+\alpha^{c}\overline{\delta}_{\mathbf{u}}^{(c)}<c_{l}$}
		    \State $c\gets c_{l}$
		\ElsIf {$c+\alpha^{c}\overline{\delta}_{\mathbf{u}}^{(c)}>c_{m}$}
		    \State $c\gets c_{m}$
		\EndIf
		\State $i\gets i+1$
		\Until{$i=I$}
	\end{algorithmic}
\end{algorithm}
\end{figure}

\begin{figure}
\centering
\includegraphics[width=\textwidth]{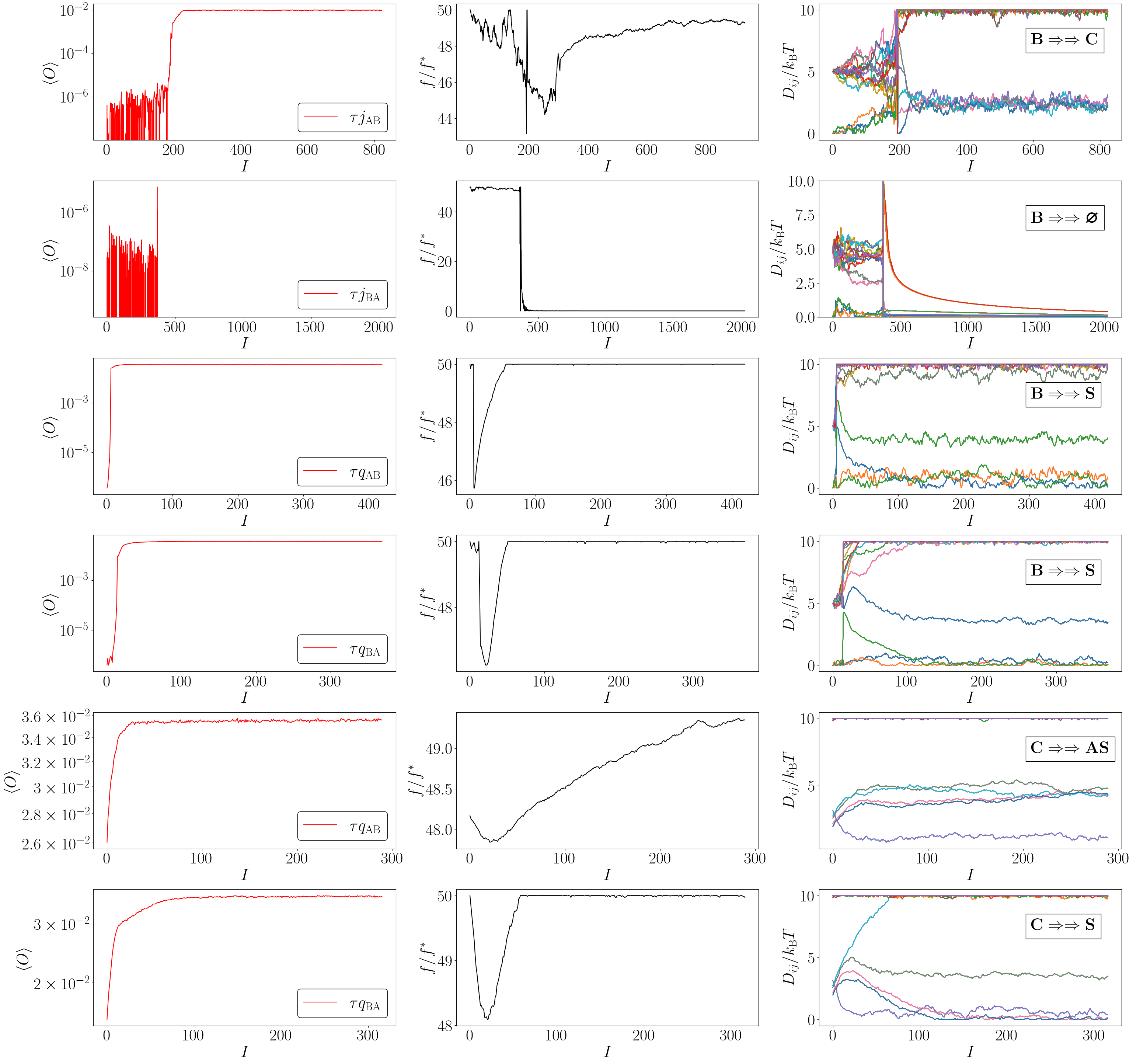}
\caption{Learning curves for conversion between polytetrahedron ($\mathrm{A}$) and octahedron ($\mathrm{B}$) as a function of optimization steps $I$. Each row shows the observable $\langle O\rangle$ being optimized on the left, $f$ optimization in the middle, and on the right the optimization of the independent $D_{ij}$ elements, $i>j$. Row-wise from top: optimization of a) $j_{\mathrm{AB}}$ starting from octahedral Maximal Alphabet $\mathbf{B}$; b) $j_{\mathrm{BA}}$ starting from $\mathbf{B}$; c) $q_{\mathrm{AB}}$ starting from $\mathbf{B}$; d) $q_{\mathrm{BA}}$ starting from $\mathbf{B}$; e) $q_{\mathrm{AB}}$ starting from $\mathbf{C}$; f) $q_{\mathrm{BA}}$ starting from $\mathbf{C}$.}
\label{fig:sifig1}
\end{figure}

We optimize the steady-state averaged cost-function $\langle\Omega\rangle$ by gradient descent, taking its explicit gradients with respect to force parameters. We choose a large positive value for $\lambda$ such that the first term is at least an order of magnitude larger than the second during the initial optimization step \cite{das2019variational}. At every optimization iteration, we fix the value of the parameters and propagate steady-state trajectories over which we estimate the gradient, using Eq. 3 in the main text. 
The temporal integration in the second term is done on-the-fly upto a time $\Delta t$. This scheme is implemented by dynamically propagating \textit{Malliavin weights} $\zeta^{(c)}(t)$ corresponding to each parameter $c$ \cite{warren2014malliavin}, as
\begin{equation}
    \zeta^{(c)}(0)=0,\qquad\dot{\zeta}^{(c)}(t)=\frac{\partial\dot{\ln}P[X]}{\partial \mathbf{u}}.\frac{\partial\mathbf{u}}{\partial c}=\frac{1}{2\gamma k_{\mathrm{B}}T}\sum_{i=1}^{N}\bm{\eta}_{i}(t)\cdot\nabla_{c}\mathbf{u}_{i}^{(c)}[\mathbf{r}^{N}(t)]
\end{equation}
where we have used Onsager-Machlup path probabilities \cite{onsager1953fluctuations} $P[X]\propto\exp\left[ -\int dt \sum_{i}(\gamma\dot{\mathbf{r}}_{i}-\mathbf{u})^{2}/4\gamma k_{\mathrm{B}}T\right] $ and $\gamma\dot{\mathbf{r}}_{i}=\mathbf{u}_{i}+\bm{\eta}_{i}$. At every time point, we store the history of Malliavin weights $\zeta^{(c)}$ upto time $\Delta t$ in the past. An efficient estimate of the gradient is then \cite{das2019variational,warren2014malliavin}
\begin{equation}
    \int_{0}^{\infty}\left\langle \delta\Omega(t)\delta\dot{\zeta}^{(c)}(0)\right\rangle  dt = \left\langle \delta\Omega(0)\delta\zeta^{(c)}(\delta t)\right\rangle -\left\langle \delta\Omega(0)\delta\zeta^{(c)}(-\Delta t)\right\rangle
\end{equation}
in the limit of large $\Delta t$, with $t=\delta t$ denoting the next timestep after $t=0$. We chose $\Delta t$ to be $5t^{*}$ for designing isolated nanoclusters and $40t^{*}$ for designing the microphase.
For the specific case of the gradient with respect to $f$, we additionally reduced its variance by accounting for the diffusion of the center of mass of the nanocluster in the large simulation box\cite{das2021variational}. With the statistical estimate, we update the parameters in the direction of the gradients and keep iterating. The learning rate $\alpha^{c}$ for every parameter $c$ is fixed at the start of the optimization by constraining the magnitude of the first optimization step to a desired value, $\Delta c_{0}$. In case that the observable changes during the optimization by several orders of magnitude, we performed the optimization first with a large learning rate, and then reduced the learning rate as soon as a large peak was detected, so that convergence remains smooth. If any optimization step takes a parameter out of the allowed range, it is reset to be at the edge of that range. The optimization is converged when the cost-function stops changing within a tolerance. A pseudocode for the optimization algorithm is provided in Algorithm \ref{algo}.

The discovery of high probability flow and current solutions is revealed in learning curves starting from different points in the alphabet space. In Figure \ref{fig:sifig1} we show learning curves for optimizing $j_{\mathrm{AB}}$, $j_{\mathrm{BA}}$, $q_{\mathrm{AB}}$ and $q_{\mathrm{BA}}$ for states $\mathrm{A}$ and $\mathrm{B}$ being the $\mathrm{C}_{\mathrm{2v}}$ and $\mathrm{O}_{\mathrm{h}}$ clusters in Figs. 1 and 2 of the main text. The starting alphabets are either the octahedral Maximal Alphabet $\mathbf{B}$ with the bond-energies set to $5k_{\mathrm{B}}T$, or the alphabet $\mathbf{C}$ obtained first from the $j_{\mathrm{AB}}$ optimization. $f/f^{*}$ is started from $50$. In all cases except the $j_{\mathrm{BA}}$ optimization, the observable is optimized, sometimes over many orders of magnitude, accompanied by a spontaneous breaking in permutation symmetry in the $D_{ij}$ parameters \cite{das2021variational}. This symmetry breaking chooses specific bond energies for specific particle indices, out of a collection of identical colloids. The optimization profiles show the stability of the symmetric flux solution $\mathbf{S}$ for high values of both $q_{\mathrm{AB}}$ and $q_{\mathrm{BA}}$. They also show the close association of the high-current solution $\mathbf{C}$ to the asymmetric high flow solution $\mathbf{AS}$, in that $\mathbf{AS}$ can be discovered by gradient descent when starting from $\mathbf{C}$.

\section*{Interpolation between optimal alphabets}

\begin{figure}
\centering
\includegraphics[width=0.7\textwidth]{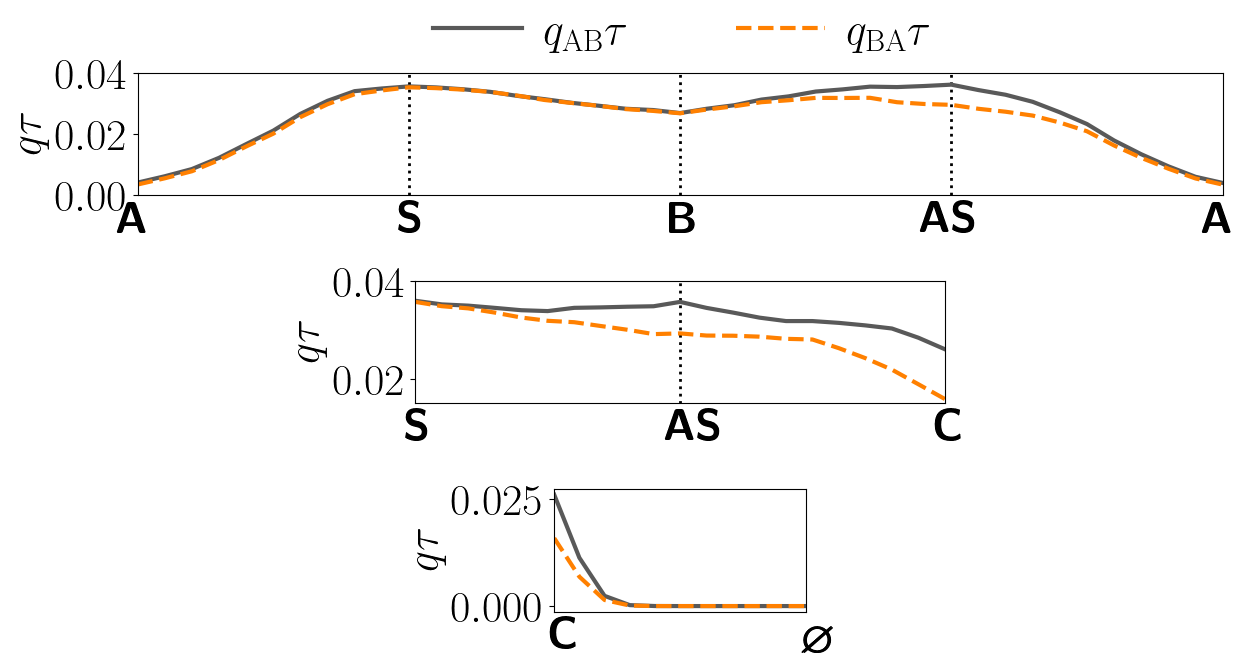}
\caption{The forward (black) and backward (orange) probability fluxes along linear interpolations in design space, keeping shear fixed at $f/f^{*}=50$.}
\label{fig:sifig2}
\end{figure}

In Fig. \ref{fig:sifig2} we have characterized the local stability of optimal designs in parameter space for maximal probability flux and current, by computing the flux along symmetry lines in design space at a fixed $f/f^{*}=50$. For comparing two different optimized alphabets, we have to first take into account the spontaneous breaking of labeling symmetry that may occur differently in the two cases. 
We choose a reference alphabet $D_{ij}^{(0)}$, which for $N=6$ and $7$ are the Maximal Alphabets for the $\mathrm{O}_{\mathrm{h}}$ cluster and one of the $\mathrm{C}_{\mathrm{3v}}$ clusters respectively. For every permutation of the particle labels $i\to p(i)$, we then compute the labeling cost-function $\Gamma[p]=\sum_{i,j|i>j}g[D^{(0)}_{ij}-D_{p(i)p(j)}]^{2}+[i-p(i)]^{2}+[j-p(j)]^{2}$, where $g$ is a scaling parameter that we choose to be $100$, and the sum is over all particles. The first term in $\Gamma$ denotes the mismatch in bond energies. If there are multiple permutations that minimize the first term, the second and third terms together consistently choose one solution that minimizes mismatch in labeling. We compute $\Gamma$ over all 720 (5040) enumerated permutations for $N=6$ ($7$) particles and choose the optimal permutation that minimizes $\Gamma$. All alphabets reported in this work have been permuted in this procedure so that they can be compared against each other and are physically meaningful.

We now plot the forward and backward flux while going from the alphabet $\mathbf{A}$ to $\mathbf{S}$ with a linear interpolation of $D_{ij}$ matrix elements, and then complete a loop by similarly visiting $\mathbf{B}$, $\mathbf{AS}$ and then $\mathbf{A}$. We find that even in this far-from-equilibrium condition, detailed balance is preserved on the line $\mathbf{A}\to\mathbf{S}\to\mathbf{B}$, and that $\mathbf{S}$ is locally optimal for fluxes in both directions. Further, the Maximal Alphabets themselves, $\mathbf{A}$ and $\mathbf{B}$, are not optimal for either high flux or for breaking detailed balance in this reaction coordinate, even in a high shear flow. Hence, our nonequilibrium design algorithm is essential to design the dynamics of assembly.

The $\mathbf{AS}$ solution breaks detailed balance and is locally optimal along this line for the forward flux but not the reverse. $\mathbf{S}$ and $\mathbf{AS}$ also appear to be in distinct solution basins along this line, but that does not imply local stability of both solutions in the entire multidimensional parameter space. 

In the second plot, we connect $\mathbf{S}$, $\mathbf{AS}$ and $\mathbf{C}$ directly through linear interpolations to study the relative stability of these strategies. We find that $\mathbf{AS}$ is only barely stable with respect to $\mathbf{S}$ for the forward flux, and almost stable with respect to $\mathbf{S}$ for the backward flux, evidenced by the flatness of the gradients of $q_{\mathrm{AB}}$ and $q_{\mathrm{BA}}$ to changing $D_{ij}$. The gap between $q_{\mathrm{AB}}$ and $q_{\mathrm{BA}}$ opened up by $\mathbf{AS}$ stays open as we interpolate to $\mathbf{C}$, which has much lower fluxes in both direction but maximizes the gap. $q_{\mathrm{AB}}$ increases $\mathbf{C}\to\mathbf{AS}$, implying there is at least one gradient descent path. Indeed, we only discovered the solution $\mathbf{AS}$ when the optimization algorithm was initiated from $\mathbf{C}$. Similarly, $q_{\mathrm{BA}}$ increases monotonically from $\mathbf{AS}$ to $\mathbf{S}$, showing why $\mathbf{AS}$ is not locally optimal for $q_{BA}$.

In the third plot, we computed the fluxes as we interpolate from $\mathbf{C}$ to $\bm{\varnothing}$, which is the trivial solution $D_{ij}=0$ and thus $Y_{\mathrm{A}}=Y_{\mathrm{B}}=q_{\mathrm{AB}}\tau=q_{\mathrm{BA}}\tau=0$. The gap between $q_{\mathrm{AB}}$ and $q_{\mathrm{BA}}$ only vanishes when going to either $\bm{\varnothing}$ or the $\mathbf{A}\to\mathbf{S}\to\mathbf{B}$ line, and never changes sign. Since the former produces a lower design cost from the Kullback-Leibler divergence term in the gradient descent cost-function $\Omega$, optimization of $j_{\mathrm{BA}}$ eventually converges to $\bm{\varnothing}$.

In summary, the symmetric high-flux solution is optimal in both directions. The asymmetric high flux solution is close to being optimal for both, and indeed in some cases for $N=7$ clusters we find distinct designs that asymmetrically amplify the flux in either direction. However, if there exists an asymmetric high flux solution but the flux is still not higher than in the reverse direction, there is no distinct high current solution.

\section*{Optimized alphabets for $N=7$}
\begin{figure}
\centering
\includegraphics[width=0.6\textwidth]{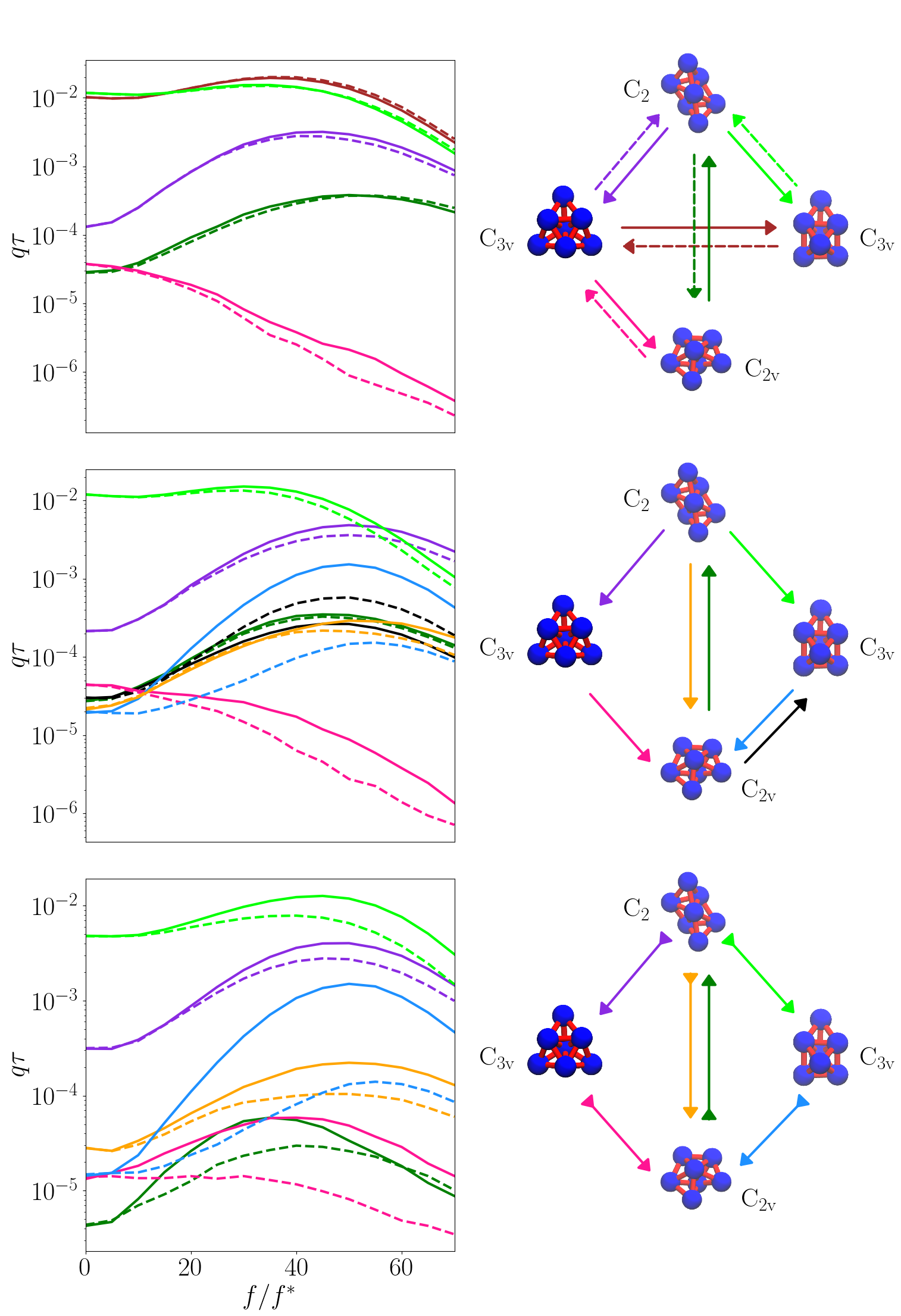}
\caption{Symmetric and asymmetric strategies to maximize probability flux and current in 7-particle clusters. (Left column) Profile of reactive fluxes with changing shear, keeping the optimized high-flux and current alphabets fixed. Top, center and bottom rows are for symmetric high-flux, asymmetric high-flux and high current solutions. Solid (dashed) lines represent forward (backward) reactive flux. Colors correspond to the edges in the reaction networks in the corresponding subfigures in the right column.}
\label{fig:n7fluxes}
\end{figure}

\begin{figure}
\centering
\includegraphics[width=\textwidth]{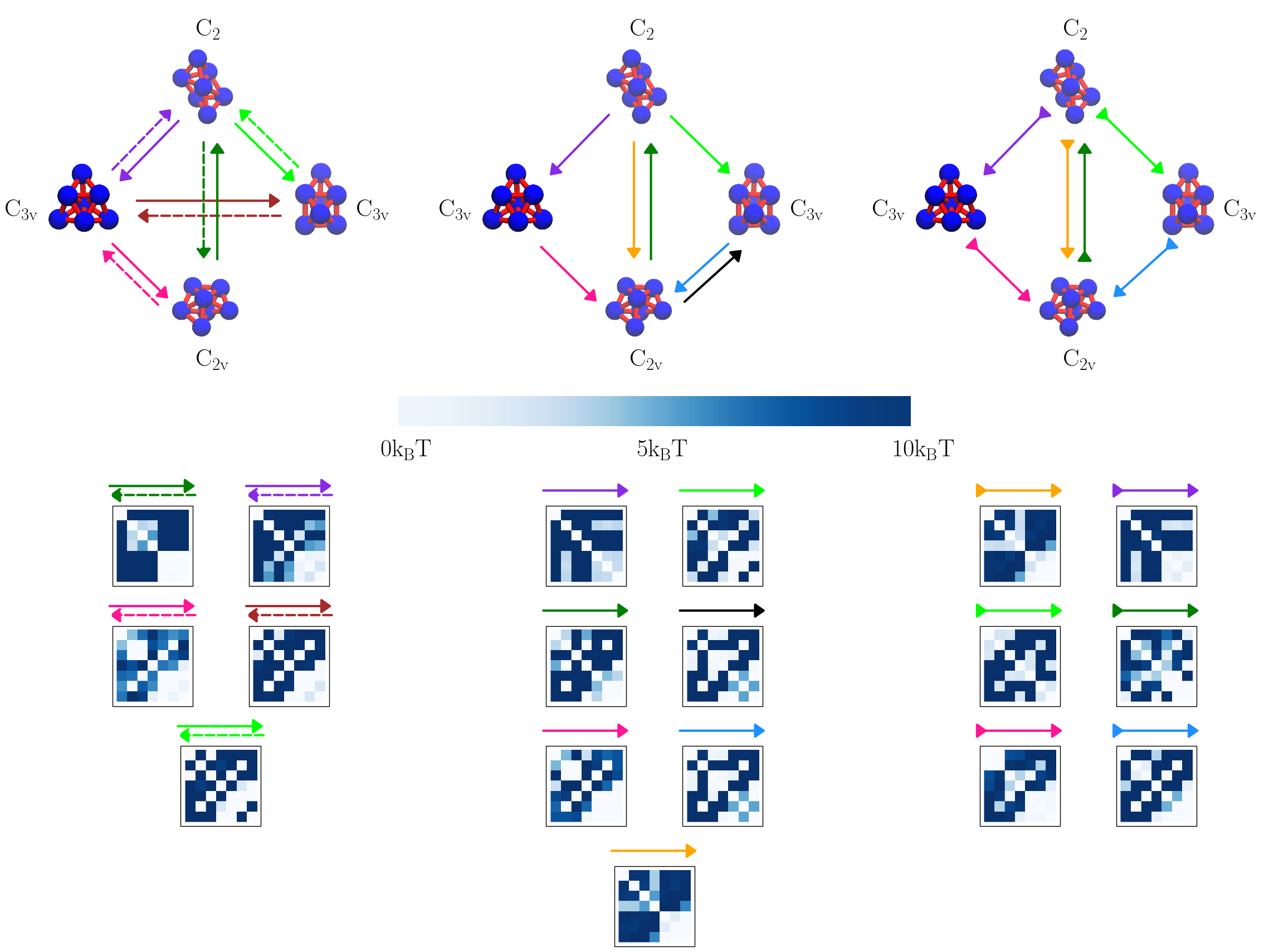}
\caption{Optimized alphabets for maximal probability flux and currents between $N=7$ clusters. The colors of the arrows in the top panel correspond to the $D_{ij}$ matrices in the bottom panel. Left column is for symmetric strategies for maximal probability flux, middle column for asymmetric strategies and right column for maximal probability current.}
\label{fig:sifig3}
\end{figure}

We show in Fig. \ref{fig:n7fluxes} and \ref{fig:sifig3} color maps of all optimized alphabets for maximizing probability fluxes and currents in the reaction network in Fig. 3 of the main text.

\section*{Optimization of hopping flux in microphase}
\begin{figure}
\centering
\includegraphics[width=\textwidth]{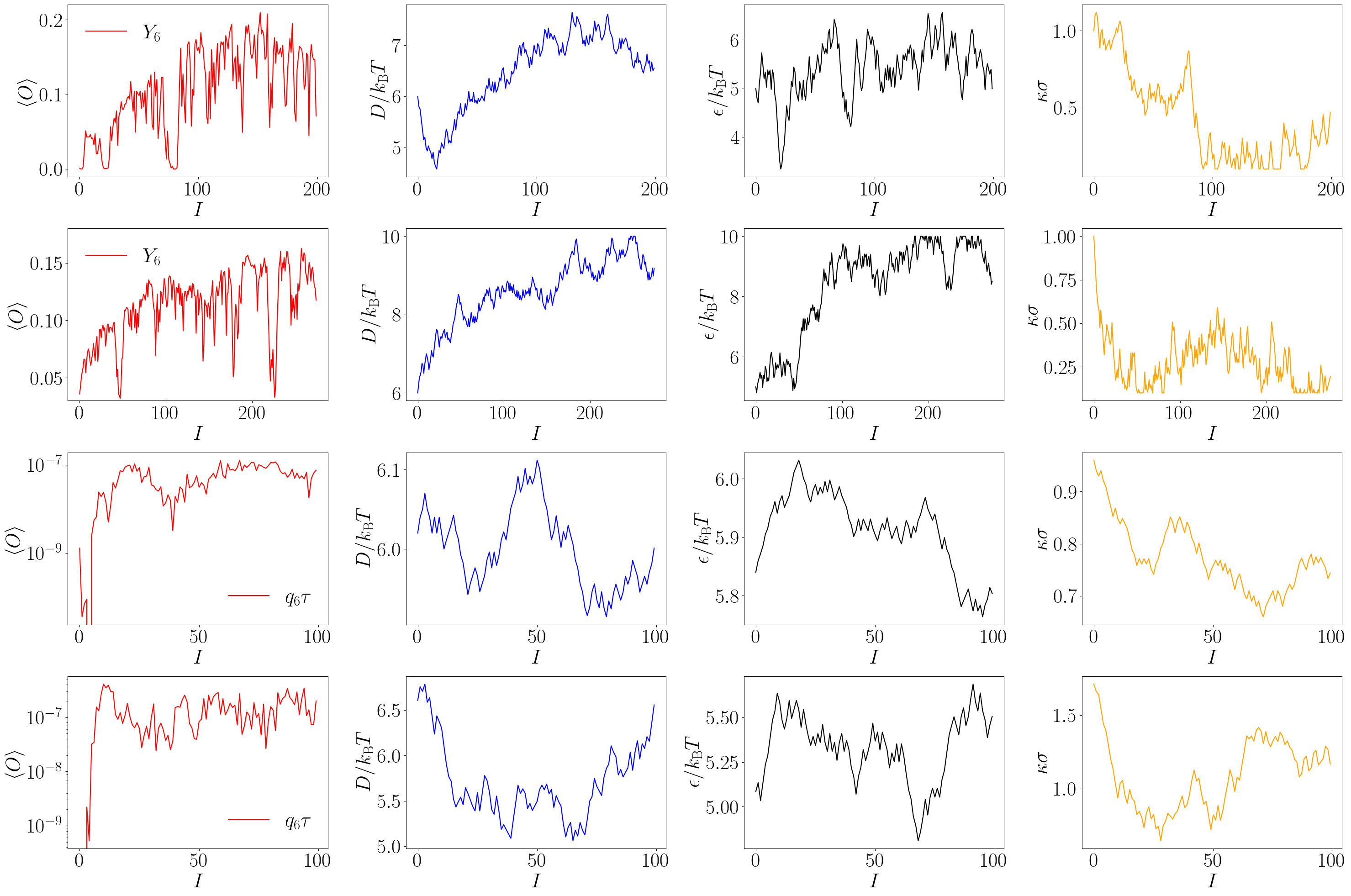}
\caption{Learning curves with the microphase for yield of compact 6-particle clusters, and hopping flux between such clusters. Each row shows the observable $\langle O\rangle$ being optimized on the left, and $D$, $\epsilon$ and $\kappa$ optimizations on the right. Row-wise from top: optimization of a) $Y_{6}$ at $f/f^{*}=1$; b) $Y_{6}$ at $f/f^{*}=25$; c) $q_{6}$ at $f/f^{*}=1$; d) $q_{6}$ at $f/f^{*}=25$.}
\label{fig:sifig4}
\end{figure}

We optimized design parameters to find maximal yield and maximum hopping flux regimes near and far from thermal equilibrium. Fig. \ref{fig:sifig4} shows the corresponding optimization curves, keeping the value of $f/f^{*}$ fixed at $1$ and $25$. As the yields and fluxes are nonzero over only small regions in the design space, the optimization needs to be started close to these regions for the observable to have measurable nonzero value over finite trajectory durations. If the observable is nonzero, the algorithm recognizes the first term in the cost-function and the observable is optimized, sometimes over multiple orders of magnitude. After convergence, the design parameters continue to explore degenerate values that all lead to the same optimum.

Fig. 4 in the main text was computed by sweeping over a rectangular grid of $D$, $\epsilon$ and $\kappa$. The grid was linearly spaced for $D$ and $\epsilon$ and logarithmically spaced for $\kappa$. We computed $\Delta E_{d}$ and $\Delta E_{a}$ for each choice of the design parameters, and coarse-grained the resultant nonuniform grid with Gaussian functions in order to interpolate and obtain the color maps in Fig. 4. The essential features of the color maps were invariant to the choice of the variance of the Gaussian functions.

\section*{Mechanism of enhanced hopping flux}

\begin{figure}
\centering
\includegraphics[width=0.7\textwidth]{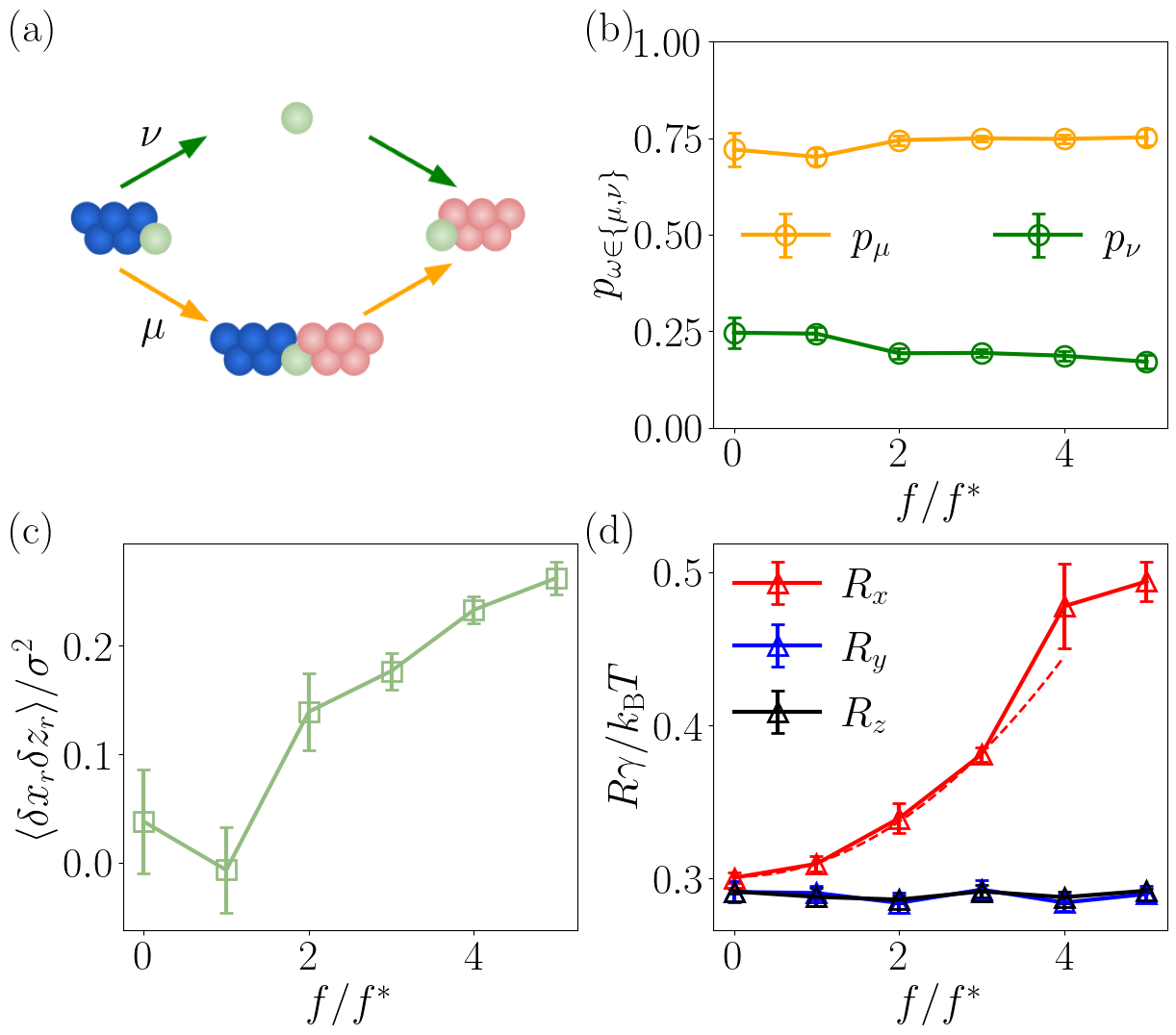}
\caption{Mechanism of enhancement in particle exchange flux with shear flow. (a) Particle exchange proceeds through two main pathways, by desorption of a particle from a cluster followed by re-adsorption (pathway $\nu$), and by the collision and subsequent fission of two clusters (pathway $\mu$). (b) Keeping energetic parameters fixed, the mechanism stays invariant within small and intermediate shear flow rates. (c) Alignment of particles conditioned to hop with respect to the center of the origin cluster, as a function of the shear flow rate. (d) Tagged-particle diffusion constants in three directions as a function of the shear flow rate. In (b), (c) and (d), solid lines are guides to the eye. In (d), dashed line shows a quadratic fit to $R_{x}$ around equilibrium.}
\label{fig:sifig5}
\end{figure}

We study the mechanism of enhancement in the particle hopping flux in the microphase due to increasing shear flow. Keeping energetic parameters fixed at the optimum for the highest near-equilibrium hopping flux (maximum point in Fig. 4d), we vary the shear flow rate and at each rate track steady-state particle exchange events between distinct 6-particle clusters with at least 9 bonds. Exchange events of duration up to $\tau=0.025t^{*}$ are counted as only they contribute to the hopping flux measured through the two-time correlation function. Tracking the exchanging particle over the duration of the hop allows us to quantify the contribution of the two possible reaction pathways, pathway $\nu$, involving the desorption of the particle and subsequent re-adsorption, and pathway $\mu$, involving the collision of two clusters, re-bonding around the hopping particle, and fission. Fig. \ref{fig:sifig5}a shows the pathways schematically, and Fig. \ref{fig:sifig5}b shows the fractional contribution of these pathways to the particle exchange reaction. We find that at equilibrium and within small to intermediate values of shear flow rate, these two are the main reaction pathways. While pathway $\nu$ could have either the desorption or the particle transport as the rate-limiting step, the rate of the main pathway $\mu$ is expected to be enhanced by faster particle and cluster transport and more frequent collisions between nanoclusters.

We find in Figs. \ref{fig:sifig5}c and d that shear flow not only enhances particle transport but also aligns clusters before the reaction, thus enhancing the overall reactive flux similar to the dimer model for nanocluster isomerization. We first measure the relative displacement from center of cluster, $(\delta x_{r},\delta y_{r},\delta z_{r})$, of a particle conditioned on starting a successful hop on the next timestep. Fig. \ref{fig:sifig5}c shows that the hopping particle in shear flow increasingly arises from an angular distribution where $\delta x_{r}$ and $\delta z_{r}$ have the same sign. This is in qualitative agreement with the steady-state angular distribution in the dimer model. Further, Fig. \ref{fig:sifig5}d shows the direction-dependent tagged-particle diffusion constant, $R_{x}\equiv \lim_{t\to\infty}\langle\Delta x (t)\rangle^{2}/2t$ and similarly for $R_{y}$ and $R_{z}$, averaged over all particles, where $\Delta x (t)$ is the (unwrapped) displacement of a tagged particle over time $t$. Around equilibrium, the diffusion constant only in the direction of the flow is amplified quadratically as expected in dilute colloidal suspensions \cite{lander2011mobility,katayama1996brownian}. The alignment of clusters and the enhancement in reaction rate through enhanced diffusion together results in a higher reactive flux for particle exchange. At far-from-equilibrium shear flow rates, when the potential energy parameters are changed to the maximum of Fig. 4f, the mechanism of flux enhancement stays the same.

%